\documentclass[prb,amssymb,showpacs]{revtex4}

\usepackage{graphicx}

\begin{document}

\title{Decoupling and Depinning II: Flux lattices in disordered
layered superconductors}
\author{Baruch Horovitz}
\affiliation {Department of Physics and Ilze Katz center for
nanotechnology, Ben-Gurion
University of the Negev, Beer-Sheva 84105, Israel}

\widetext
\begin{abstract}
Phase transitions of a flux lattice in layered superconductors
with magnetic field perpendicular to the layers and in presence of
disorder are studied. We find that disorder generates a random
Josephson coupling between layers which leads to a Josephson glass
(JG) phase at low temperatures; vanishing of the JG order
identifies a depinning transition. We also find that disorder and
thermal fluctuations lead to layer decoupling where the
renormalized Josephson coupling vanishes. Near decoupling an
anharmonic regime is found, where usual elasticity and the
resulting Bragg glass are not valid. The depinning line crosses
the decoupling line at a multicritical point, resulting in four
transition lines. The phase diagram is
consistent with the unusual data on $Bi_2Sr_2CaCu_2O_8$ such as
the "second peak" and depinning transitions. The Josephson plasma
frequency is evaluated in the various phases.

\end{abstract}

\maketitle

\newcommand{\rr}{\mbox{${\bf r}$}}
\newcommand{\RR}{\mbox{${\bf R}$}}
\newcommand{\QQ}{\mbox{${\bf Q}$}}
\newcommand{\uu}{\mbox{${\bf u}$}}
\newcommand{\qq}{\mbox{${\bf q}$}}
\newcommand{\half}{\mbox{$\frac{1}{2}$}}
\newcommand{\rhob}{\mbox{${\mbox{\boldmath $\rho$}}$}}
\newcommand{\lam}{\mbox{${\lambda_{ab}}$}}

\section{Introduction}
The phase diagram of layered superconductors in a magnetic field
$B$ perpendicular to the layers is of considerable interest in
view of extensive experiments on high temperature superconductors
\cite{Kes}. A first order transition in $YBa_{2}Cu_{3}O_{7}$
(YBCO) and in $Bi_2Sr_2CaCu_2O_8$ (BSCCO) has been interpreted as
a melting transition of the flux lattice. The data suggests that
the first order line terminates at a multicritical point, which
for BSCCO \cite{Khaykovich1,Khaykovich2} is at $B_0\approx
300-10^3 G$ and $T_0\approx 40-50 K$, while for YBCO
\cite{Deligiannis} it is at $B_0\approx 2-10 T$ and $T_0\approx
60-80 K$, depending on disorder and oxygen concentration. The
multicritical point also terminates a "second peak" transition
\cite{Kes,Khaykovich1,Khaykovich2,Deligiannis} which is manifested
by a sharp increase in magnetization and in critical current. The
transition line at $B\approx B_0$ and $T<T_0$ is weakly $T$
dependent and was found, for BSCCO, to be smoothly connected with
the first order line \cite{Avraham}. Neutron scattering and
$\mu$SR data \cite{Kes,Cubitt} show that positional correlations
of the flux lattice are significantly reduced near these phase
boundaries, except however, near the multicritical point where a
reentrant behavior is observed \cite{Forgan}. Data on
$Nd_{1.85}Ce_{0.15}CuO_{4-\delta}$ (NCCO) has also shown a second
peak transition; here, however, $B_0$ decreases with temperature
near the superconducting transition at $T_c\approx 23 K$ with no
apparent multicritical point \cite{Giller}. The second peak
phenomena is also pronounced in other layered systems such as
$NbSe_2$ \cite{Higgins,Marley} and in $Pb/Ge$ multilayers
\cite{Bruynseraede}. Recent decoration data \cite{Fasano} on
$NbSe_2$ has shown that the topology of the vortex structure is
 weakly affected by crossing part of the second peak line. Hence the nature
 of the phase at $B>B_0$ is not well established.

The Josephson plasma resonance is a probe of the Josephson
coupling \cite{Matsuda,Koshelev} and can be used to probe the
various phase transitions. Recent data on BSCCO has indeed shown a
significant reduction in the resonance frequency at the second
peak transition \cite{Shibauchi,Matsuda2}.

In a remarkable experiment Fuchs et al. \cite{Fuchs} have
shown that the phase diagram of BSSCO is much more elaborate. They
show that the spatial distribution of an external
 current exhibits a transition from bulk pinning to surface pinning of
 vortices with most of the current flowing at the sample edges.
This depinning line crosses the multicritical point and its
temperature
is almost $B$ independent at $B<B_0$. The depinning
transition correlates with anomalies in vibrating reed experiments
\cite {Kopelevich} and in magnetization \cite{Dewhurst}.
 Thus there are four transition
lines which emanate from the multicritical point at $B_0$, $T_0$:
The first order line, the second peak line and depinning lines for
both $B<B_0$ and $B>B_0$. The common intersection of these four transition lines was also
seen in data of the c axis Josephson critical current \cite{Ooi}.
This critical current decreases significantly above the second peak line (in
contrast with the critical current parallel to the layers) and
also decreases in the depinned regimes.

 The notion of vortex matter in the presence of disorder has
emerged as a fundamental problem of elastic manifolds in a random
media \cite{Blatter}.  This has motivated
an extensive theoretical effort towards understanding the
field-temperature ($B-T$) phase diagram in presence of disorder.
Impurity disorder does not allow long range
translational order of the flux lattice and finite domains are
expected \cite{LO}. At low temperatures and fields the system is a
Bragg glass \cite{Giamarchi,Natterman}, i.e. the lattice is
dislocation free, at long scales the displacement correlations
decay as a power law and Bragg peaks are expected. The impurity
induced domains are essential for the description of both
equilibrium, e.g. thermodynamic phase transitions and
non-equilibrium, e.g. critical current phenomena.
 Melting, e.g.,  is expected to occur by
thermal or disorder induced dislocations, as indeed demonstrated
for fields parallel to the layers \cite{Carpentier,Golub}.
Numerical simulations on related XY models have also shown
disorder induced melting \cite{Olson1,Nonomura,Olson2}.

   The flux lattice can undergo a transition which is unique to
   layered superconductors, i.e.  a decoupling transition
\cite{Glazman1,Daemen}. In this transition the Josephson coupling
between layers vanishes while the lattice is maintained by the
electro-magnetic coupling between layers.
 A disorder induced decoupling was also proposed
as a crossover phenomena \cite{Koshelev1}. Decoupling in presence
of columnar defects was also studied \cite{Morozov}, showing
enhancement of the coupled phase.

It has been shown that decoupling coalesces with a defect
unbinding transition \cite{Dodgson,ledou} which has analogs in
isotropic systems \cite{Frey}. The resulting vacancies and
interstitials lead to a reduction in the elastic tilt modulus
\cite{Marchetti}, consistent with the decoupling scenario as
described below.
 It is possible then that a decoupling-defect
transition accounts for the peak phenomena in all type II
superconductors.
The analysis below is, however, presented for layered anisotropic
systems
where quantitative predictions can be made. Vacancies and
interstitials are neglected; their role is dicussed in the concluding
section of the preceding companion article \cite{GH}.

   In the present work we expand our previous work \cite{HG,H1} and
study effects of disorder at temperatures below the melting
temperature $T_m$ by employing replica symmetry breaking (RSB)
methods. The RSB methods are accurate when couplings of the nonlinear terms are weak. E.g. in the pure case they  reproduce the RG result at weak Josephson coupling \cite{GH}; in the related problem of vacancies and interstitials it was shown that RSB accurately locates a disorder induced transition \cite{ledou}.  In the present problem weak coupling corresponds to weak Josephson coupling and weak disorder. Weak disorder can be stated as a condition on the size of domains $R_{BG}$ being larger than the renormalized penetration length in the $c$ direction. This condition is examined in section IV and the RSB actually detects this by producing a stronger singularity (appendices A-C). 
Furthermore, RSB as a variational method can identify order parameters and determine the form of the phase diagram. The critical behavior near the transition, however, is not expected to be accurate.

The most interesting finding in this work is that of a glass order
parameter which we term as Josephson glass (JG), as it is due to
disorder induced on the Josephson coupling. The JG order is
expected to lead to stronger pinning, hence the line where JG
vanishes is associated with a depinning line. We find that the JG
and decoupling lines cross and lead to four distinct phases which
meet at one point in the $B-T$ phase diagram, remarkably close to
the experimental phase diagram \cite{Fuchs,Ooi}. This paper follows a
companion one \cite{GH} where the decoupling transition is studied
in the pure system by second order renormalization group (RG).

The full problem addressed here involves the following set of
nonlinearities: (i) Josephson coupling which involves both pancake
displacements and a nonsingular phase. (ii) A disordered Josephson
coupling which leads to the JG order. (iii) A nonlinear coupling
of disorder to the displacement pattern, leading to the well
studied Bragg glass (BG) \cite{Giamarchi,Natterman}. After
presenting the model in section II,  we study in section III a
simplified version of the full problem in which the nonsingular
phase is neglected and also the disorder coupling is linearized,
corresponding to scales within finite domains. These
approximations lead to an unphysical divergence of an integral
$I(z)$ where $z$ is the renormalized Josephson coupling, i.e.
$z\rightarrow 0$ at decoupling. In section III we assume that
$I(z)$ is convergent and behaves as $\sim \ln z$, an assumption
that is justified in appendices A, B and C. In appendix A we
extend section III to solve the combined BG/JG system, though the
nonsingular phase is neglected. In appendix B the BG system
including the non-singular phase is solved, but JG is neglected,
as relevant to thermal decoupling. In both appendices A and B we
find an additional $\ln ^2z$ term which signals a divergence of
disorder effects in a regime close to decoupling. In appendix C we
study JG with the nonsingular phase, but disorder is linearized.
It is shown that $I(z)$ converges even in this situation, while an
additional $1/\sqrt{z}$ term is generated.
 In section IV we present a dimensional
derivation of domain sizes which correctly reproduces the pinning
and BG lengths. Near decoupling there is a regime of nonlinear
elasticity with an apparent jump of the tilt modulus $c_{44}$ and
the critical current. This anharmonic regime coincides with the
onset of the $\ln^2z$ term in appendices A, B. In section V the
Josephson plasma frequency is studied, being an efficient probe
for identifying the various phases. In section VI we discuss
available data on the second peak and depinning transitions. We
propose that decoupling accounts for the main features of the
second peak transition while the depinning transitions correspond
to the onset of JG order.

\section{The Model}
   Consider a flux lattice with an equilibrium position of the $l$-th
flux line at vectors ${\bf R}_{l}$ of a regular two-dimensional
lattice. The flux line is composed of a sequence of singular
points, or "pancake" vortices, whose positions at the $n$-th layer
can fluctuate to ${\bf R}_{l} + {\bf u}_{l}^{n}$. Of particular
interest is the transverse part of ${\bf u}_{l}^{n}$ with the
Fourier transform
 $u_T({\bf q},k)$, where ${\bf q}, k$ are
wavevectors parallel and perpendicular to the layers,
respectively. The elastic energy due to the electromagnetic
coupling has the form
\begin {equation}
{\cal H}_{e-m} = \half \sum_{{\bf q},k} (da^{2})^{2}[c_{66}^{0}q^{2} +
c_{44}^{0}(k)k_{z}^{2}] |u_{T}({\bf q},k)|^{2}
\end{equation}
where the flux line density is $1/a^{2}$, $d$ is the spacing
between layers, ${\bf q}$ is within the Brillouin zone [of area
$(2\pi /a)^{2}$], $|k|<\pi /d$ and $k_{z}=(2/d)\sin (kd/2)$. The
shear and tilt moduli are given (for $a\gg d$) by
\cite{GK,Sudbo,GH2}
\begin{eqnarray}
c_{66}^{0} & = & \tau/(16da^{2})  \nonumber\\
c_{44}^{0}(k) & = & [\tau/(8da^{2}\lambda _{ab}^{2}k_{z}^{2})] \ln
(1+a^{2}k_{z}^{2}/4\pi)
\end{eqnarray}
where $\tau = \phi_{0}^{2}d/(4\pi ^{2}\lambda_{ab}^{2})$ sets the
energy scale and $\lambda_{ab}$ is the magnetic penetration length
parallel to the layers; $\tau \approx 10^3 -10^4 K$ for YBCO or
BSCCO parameters \cite{Kes}. Note the strong dispersion of
$c_{44}^{0}(k)$ so that $c_{44}^{0}(k)$ decreases by the large
factor $(d/a)^{2}$ when $k$ varies from $k\lesssim 1/a$ to
$1/a\lesssim k<\pi /d$.

The Josephson phase between the layers $n$ and $n+1$ at position
${\bf r}$ in both layers involves contributions from a nonsingular
component
 and from singular vortex terms. The singular
phase around a pancake vortex at position ${\bf R}_{l} + {\bf
u}_{l}^{n}$ is $\alpha ({\bf r} - {\bf R}_{l} - {\bf u}_{l}^{n})$
where $\alpha ({\bf r}) = \arctan (y/x)$ with ${\bf r} = (x,y)$.
We assume that all vortices belong to the flux lines, i.e. there
are no free vacancies or interstitials.

 The Josephson phase
involves the interlayer phase difference from the pancake
singularities $\alpha ({\bf r} - {\bf R}_{l} - {\bf u}_{l}^{n}) -
\alpha ({\bf r} - {\bf R}_{l} - {\bf u}_{l}^{n+1})$, which after
expansion in ${\bf u}_{l}^n$ becomes (Eq. 19 of the companion
article \cite{GH})
\begin{equation}\label{b}
b_{n}(\rr)=2\pi i d \int_{BZ}\frac{d^{2}\qq d k}{(2\pi)^{3}}e^{-i
\qq \cdot \rr-i k n d} (e^{i k d}-1)\frac{u_T(\qq,k)}{\qq}\,.
\end{equation}

We consider first a simplified model which neglects the
nonsingular part of the Josephson phase. The nonsingular phase is
essential for evaluating displacement fluctuations (section IV),
however for the purpose of the phase transitions under study it
can be neglected (justified by Appendices A,B). We have then an
effective Hamiltonian for wavevectors $|q|<Q_0$, (Eq. (23) of the
companion article \cite{GH})
\begin{equation}\label{H}
{\cal H}_{pure}^{(1)}/T =\half\sum_{\qq,k}c(q,k)q^2 |b(\qq,k)|^{2}
-\frac{E_J}{T}\sum_{n}\int d^{2}\rr \cos b_{n}(\rr) \label{Hpure}
\end{equation}
where $E_J$ is the interlayer Josephson coupling energy per unit area
and
\begin{equation}\label{cqk}
c(q,k)=\frac{a^4}{(2\pi
d)^2T}[c_{44}^0(k)+\frac{q^2}{k_z^2}c_{66}^0]\equiv
c(k)+c'\frac{q^2}{k_z^2}
\end{equation}
The last equality defines $c(k)$ and $c'$, i.e.
\begin{eqnarray}\label{cc'}
c(k)&=&\frac{\tau a^2}{32\pi^2 Td^3\lam^2}\frac{\ln
(1+a^2k_z^2/4\pi)}{k_z^2}\nonumber\\
c'&=&\frac{\tau a^2}{64\pi^2Td^3}\,.
\end{eqnarray}
Since ${\bf \nabla} \alpha \sim 1/r$ decays slowly, even if ${\bf
u}_{l}^{n}$ are small the contribution of many vortices which move
in phase ($q\rightarrow 0$) leads to a divergent response of
$b_{n}({\bf r})$, i.e the $1/q$ factor in Eq. (\ref {b}). This
leads to a decoupling transition \cite{Daemen,HG,GH}, which at
weak $E_J$ is (Eqs. (27, 40) of the companion article \cite{GH})
\begin {equation}
T_{d}^0=\frac {4a^{4}}{d^{2}}(\int \frac{dk}{c_{44}^0(k)})^{-1}
\approx \frac{\tau a^{2}\log (a/d)}{4\pi \lambda_{ab}^{2}} \,.
\label{Td}\end{equation}

We note that melting and related dislocations have been neglected.
An estimate of $T_m$ by the Lindemann criterion yields
\cite{Blatter,GH2} $T_m\approx \tau$, hence our description near
$T_d^0$ is limited to to $a \lesssim \lambda_{ab}$. Melting in the
absence of Josephson coupling was in fact studied \cite{melting},
showing that $T_m$ is between $\tau/8$ and the two-dimensional
melting temperature of $\approx 0.004\tau$, approaching the latter
at high fields $a\ll \lambda_{ab}$. At intermediate fields the present description is then valid at $a\lesssim 0.4\lambda_{ab}$. 
However, for disorder induced melting we estimate (see the discussion section VI) that 
for BSCCO parameters the decoupling field is below the melting field if 
$a\gtrsim 0.14\lambda_{ab}$, consistent with the low temperature second peak field value.

We proceed now to study the disorder term. A second assumption of
the simplified version is that of linearized disorder, i.e. small
fluctuations $|{\bf u}_{l}^{n}|\ll a$. Consider a short range
pinning potential $U_{pin}^{n}({\bf r})$ with the coupling
\begin{equation}\label{Hpin1}
{\cal H}_{pin}=\int d^{2}r \sum_{n,l} U_{pin}^{n}({\bf r}) p({\bf
r} - {\bf R}_{l} - {\bf u}_{l}^{n})
\end{equation}
where $p(\rr)$ is a shape function for a vortex of size $\xi_0$ and
the disorder has short range correlation
\begin{equation}\label{Hpin}
\langle U_{pin}^n(\rr)U_{pin}^{n'}(\rr')\rangle=\half{\bar U}
\delta_{n,n'}\delta (\rr-\rr')
\end{equation}
Expanding Eq. (\ref {Hpin1}) to first order in ${\bf u}_{l}^{n}$
and averaging $U_{pin}^{n}({\bf r})$ by the replica method
\cite{Mezard,Giamarchi} leads to a disorder term in the free
energy
\begin{equation}\label{Hdis}
{\cal H}_{dis}^{(1)}/T=\frac{{\bar U}{\bar p}}{4T^2}\sum_{n.l}
\sum_{\alpha,\beta}\uu_l^{n,\alpha}\cdot\uu_l^{n,\beta}
\end{equation}
where $\alpha ,\beta$ are replica indices. The average involves
\begin{equation}
\int \partial_ip(\rr)\partial_jp(\rr)d^2r=\bar{p}\delta_{ij}
\end{equation}
with $\bar{p}$ of order 1.

The replicated Hamiltonian of the simplified version, keeping only
transverse displacements, is therefore
\begin {eqnarray}
{\cal H}^{(1)}/T&=&\frac {1}{2}\sum_{{\bf q},k;\alpha
,\beta}[c(q,k)q^{2}\delta_{\alpha
,\beta}-s_{0}\frac{q^{2}}{k_{z}^{2}}]
b^{\alpha}({\bf q},k)b^{\beta *}({\bf q},k)   \nonumber \\
& & -\frac{E_J}{T}\sum_{n;\alpha}\int d^{2}r\, \cos
b_{n}^{\alpha}({\bf r}) - \frac{E_v}{T}\sum_{n;\alpha\neq \beta}
\int d^{2}r \cos [b_{n}^{\alpha}({\bf r}) - b_{n}^{\beta}({\bf r})]
\label{Hreplica}\end{eqnarray}
where
\begin{equation}
s_{0}=\frac{\bar{U}\bar{p}}{2T^2}\frac{a^{2}d}{(2\pi d^{2})^2}\,.
\end{equation}
It is found useful below to define a dimensionless disorder
parameter $s$,
\begin {equation}\label{s}
s=\frac{8\pi {\bar U}{\bar p}\lambda_{ab}^4}{\tau ^2a^2\ln ^2
(a/d)}\, .
\end {equation}
 The inter-replica Josephson coupling, i.e. the $E_v$ term in
Eq.~(\ref{Hreplica}),
 is generated from the $E_J$ term in second order renormalization
group (RG). It is
 essential to keep the $E_v$ term from the start since it couples
 different replica indices and leads to distinct physics by RSB, as
shown below. Physically, the $E_v$ term originates from random 
displacements of pancake vortices due to intralayer 
impurities. The pancake vortices are then not one on top of the other, resulting in random segments 
of Josephson vortices, i.e. vortices parallel to the layers. The latter represents random Josephson phases, whose replica average leads to the $E_v$ term in Eq. (\ref{Hreplica}).

We proceed now to present the full model, which extends Eq.
(\ref{Hreplica}) to include the nonsingular phase as well as
nonlinear disorder. The Josephson phase involves a nonsingular
phase $\theta_n(\rr)$ in addition to the pancake fluctuations via
$b_n(\rr)$. The Hamiltonian of the pure system is then (Eq. 21 of
the companion article \cite{GH})
\begin {eqnarray}\label{Hpure2}
{\cal H}_{pure}\{b,\theta\}/T =&& \half \sum_{q,k} G_{f}^{-1}({\bf
q},k) |\theta({\bf q},k)|^{2}  + \half \sum_{{\bf q},k} c({\bf
q},k)q^2|b({\bf q},k)|^{2}\nonumber\\
 &&-\frac{E_J}{T} \sum_n \int
d^{2}r \, \cos [\theta_n({\bf r}) + b_{n}({\bf r})]
\end{eqnarray}
where
\begin {equation}
G_{f}(q,k)=\frac{4\pi d^3T}{\tau q^2}(\lambda_{ab}^{-2}+k_z^2)\,.
\end{equation}

Consider next the general form of the disorder coupling
\cite{Blatter}. Using the relation
\mbox{$\sum_l\delta^2(\rhob-\RR_l)=\sum_le^{i\QQ_l\cdot \rhob}$}
where $\QQ_l$ are reciprocal lattice vectors, the disorder
coupling (\ref {Hpin1}) becomes
\begin{equation}
{\cal H}_{pin}=-\int d^2r \sum_n U_{pin}^n(\rr)\int
\frac{d^2\rho}{a^2}p[r-\rhob-\uu^n(\rhob)]\sum_le^{i\QQ_l\cdot
\rhob}
\end{equation}
For $|\QQ_l|<1/\xi_0$ we can replace $p(\rr)$ by
$\xi_0^2\delta^2(\rr)$ so that
\begin{equation}\label{Hpin2}
{\cal H}_{pin}=-\frac{\xi_0^2}{a^2}\int d^2r \sum_n
U_{pin}^n(\rr)[1+{\mbox{\boldmath
$\nabla$}}\cdot \uu^n(\rr)]^{-1}\sum_le^{i\QQ_l\cdot (\rr-\uu^n(\rr))}
\end{equation}
The coupling to long wavelength modes via ${\mbox{\boldmath
$\nabla$}}\uu^n(\rr)$ is irrelevant \cite{Giamarchi} in 3D so that
the replica average of ${\cal H}_{pin}$ becomes
\begin{equation}
{\cal H}_{dis}/T=\frac{g_0}{a^2}\sum_{\QQ,\alpha,\beta,n}\int d^2r
\cos [\QQ\cdot (\uu^{n,\alpha}(\rr)-\uu^{n,\beta}(\rr))]
\end{equation}
with $g_0={\bar U}\xi_0^4/T^2a^2$. To relate this form to the
linearized one (\ref {Hdis}) we expand in $\uu^{n,\alpha}(\rr)$
and use $\sum_{\QQ}\QQ^2\approx
\frac{a^2}{4\pi}\int^{1/\xi_0^2}Q^2dQ^2\approx a^2/8\pi\xi_0^4$ so
that (\ref {Hdis}) is obtained if ${\bar p}\approx 1/2\pi$. The
coupling $g_0$ can then be written as
\begin{equation}\label{g}
g_0=\frac{{\bar U}\xi_0^4}{T^2a^2}=s\frac{\tau^2\xi_0^4\ln^2
(a/d)}{4T^2\lam^4}\,.
\end{equation}
We are interested here in BG effects on the $q\rightarrow 0$
singularity associated with the decoupling transition, i.e. the
long range properties of the BG. The BG domain size is defined by
the scale $R$ where the displacement correlation starts to diverge
as $\ln r$. It is reasonable to expect that this scale is
determined by the shortest $\QQ$, as indeed shown for a system
with regular elasticity \cite{Giamarchi}, i.e. far from
decoupling. We consider then the disorder term with just the
shortest reciprocal wavevectors $|\QQ|\approx 2\pi/a$ (e.g. six
wavevectors in the hexagonal lattice). The full Hamiltonian is
then
\begin{eqnarray}\label{Hfull}
{\cal H}/T=&& \sum_{\alpha}\frac{1}{T}{\cal
H}_{pure}\{b^{\alpha},\theta^{\alpha}\}-
\frac{E_v}{T}\sum_{n;\alpha\neq \beta}\int d^{2}r \cos
[b_{n}^{\alpha}(\rr) - b_{n}^{\beta}(\rr)+
\theta_{n}^{\alpha}(\rr) - \theta_{n}^{\beta}(\rr)]\nonumber\\
&&-\frac{g_0}{a^2}\sum_{\alpha\neq\beta,n}\int d^2r \cos [\QQ\cdot
(\uu^{n,\alpha}(\rr)-\uu^{n,\beta}(\rr))] \,.
\end{eqnarray}

We note finally that a similar two-dimensional (2D) model has been
studied by RSB and RG methods \cite{H3,Scheidl}. As shown in the
next section, finite values of $k$ dominate the phase transitions,
so that a certain $k$ averages of the coefficients in Eq. (\ref
{Hreplica}) lead to a 2D problem with the same $q$ singularities
as in (\ref {Hreplica}). Indeed the RSB solution below has the
same structure as the 2D case \cite{H3} with a temperature
parameter $t=T/T_d^0$ and a disorder parameter $s$ (Eq. {\ref
{s}). In view of this similarity, it is useful to quote the RG
equations of the 2D model \cite{H3} in terms of $u=\xi^2E_J/T$ and
$v=\xi^2E_v/T$,
\begin{eqnarray} \label{RG}
d\,u&=&[2u(1-t-s) -2\gamma' uvt]d \,\ln \xi \nonumber\\
d\,v&=&[2v(1-2t)+\half \gamma' su^2-2\gamma' tv^2]d\,\ln \xi
\nonumber\\
d\,t&=&-2\gamma''^2(t+s)t^2u^2d\,\ln \xi \nonumber\\
d\,(s/t^2)&=&16\gamma''^2tv^2d\,\ln \xi
\end{eqnarray}
where the initial value of the scale $\xi$ is $a$ and
$\gamma',\gamma''$ are numbers of order 1. We quote these results
so that the necessity of the $E_v$ term is shown more concretely.
Indeed $E_v$ is generated by $sE_J^2$ while at $t<\frac{1}{2}$ it
is relevant on its own. Furthermore, the RG results will be used
to qualitatively support and supplement the phase diagram, as
derived by RSB in the next section.

\section{Phase Diagram}

In this section we consider the simplified version, Eq. (\ref
{Hreplica}). This assumes that displacements are within finite
domains and Bragg glass effects are neglected; also the
nonsingular phase is neglected here. Appendices A and B show that
these assumptions are justified for the purpose of our phase
diagram. The nonsingular phase is essential for evaluating
displacement fluctuations, as studied in Appendix C.

We proceed by using the RSB method \cite{Mezard}. 
The RSB method proceeds by employing a variational
free energy ${\cal F}_{var}={\cal F}_{0}+<{\cal H}-{\cal H}_{0}>$
with ${\cal F}_{0}$ the free energy corresponding to
 \begin{equation}\label{H0}
{\cal H}_{0}=\half\sum_{{\bf q},k;\alpha ,\beta}G_{\alpha ,\beta}^{-1}
({\bf q},k) b^{\alpha}({\bf q},k)b^{\beta *}({\bf q},k)
\end{equation}
 and $G_{\alpha ,\beta}({\bf q},k)$ is determined by an extremum
condition
on $F_{var}$. We define the following averages $\langle ...\rangle_0$
with respect to $H_0$,
\begin{eqnarray}
\langle \cos b_n^{\alpha}(\rr)\rangle_0&=&e^{-\half
A_{\alpha}}\nonumber\\
A_{\alpha}&=&\sum_{{\bf q},k} G_{\alpha,\alpha}(q,k)\label{Aa}\\
\langle \cos[ b_n^{\alpha}(\rr)-b_n^{\beta}(\rr)\rangle_0&=&
e^{-\half B_{\alpha,\beta}}\nonumber\\
B_{\alpha ,\beta}&=&2\sum_{{\bf q},k}[G_{\alpha ,\alpha}({\bf q},k)
-G_{\alpha ,\beta}({\bf q},k)]\label{Bab}
\end{eqnarray}
so that
\begin{eqnarray}\label{Fvar}
{\cal F}_{var}/T=&&\half\sum_{\qq,k}Tr  [\ln G(q,k)
+(G^{-1}(q,k)-c(q,k)q^2{\hat I}- s_0\frac{q^2}{k_z^2}{\hat
L})G(q,k)]\nonumber\\
&&-\frac{E_J}{T}\sum_{\alpha}e^{-\half A_{\alpha}}-
\frac{E_v}{T}\sum_{\alpha \neq \beta}e^{-\half B_{\alpha,\beta}}
\end{eqnarray}
where ${\hat I}_{\alpha,\beta}=\delta_{\alpha,\beta}$ and ${\hat
L}_{\alpha,\beta}=1$.

The variational equation $\delta F_{var}/\delta G_{\alpha,\beta}=0$
yields
\begin{eqnarray}
G_{\alpha ,\beta}^{-1}(q,k)&=&[c(q,k)q^{2}+z]\delta_{\alpha ,\beta}
-s_{0}(q^{2}/k_{z}^{2})-\sigma_{\alpha ,\beta}   \label{extremuma} \\
z &=& \frac{E_J}{Td}e^{-\half A_{\alpha}}   \label{extremumb} \\
\sigma_{\alpha ,\beta} &=& \frac{E_v}{Td}[e^{-\half B_{\alpha
,\beta}} -\delta_{\alpha ,\beta}\sum_{\gamma}e^{-\half B_{\alpha
,\gamma}}] \label{extremumc}
\end{eqnarray}
where $z$ is a renormalized Josephson coupling. In the replica
limit with the number of replicas $n\rightarrow 0$ the RSB
\cite{Mezard} method represents each matrix as a hierarchy of
matrices, e.g. $\sigma_{\alpha ,\beta}$ is represented by
$\sigma(u)$, with $0<u<1$ and a diagonal component
$\tilde{\sigma}$. We parameterize therefore $G_{\alpha
,\beta}^{-1}$ by $\tilde{a}$ and $a(u)$, where
\begin{eqnarray}
\tilde{a}&=&c(q,k)q^2-s_0\frac{q^2}{k_z^2}+z-\tilde{\sigma}\nonumber\\
a(u)&=& - s_0\frac{q^2}{k_z^2}-\sigma(u)\,.
\end{eqnarray}
The amount by which the replica symmetry is
broken is measured by a glass order parameter
\begin{equation}\label{Du}
\Delta (u)=u\sigma (u)- \int_{0}^{u}\sigma(v)\,dv \,.
\end{equation}
 The inverse matrix $G_{\alpha ,\beta}$ is represented by $\tilde
{b}$ and $b(u)$, where \cite{Mezard} (see also Appendix B of Ref.
\onlinecite{H3})
\begin{eqnarray}\label{inversion}
\tilde {b}&=&\frac{1}{\tilde{a}-\langle a \rangle}\left
[\frac{-a(0)}{\tilde{a}-\langle a \rangle}+1+
\int_0^1\frac{dv}{v^2}\frac{\Delta(v)}{\tilde{a}-\langle a \rangle
+\Delta(v)}\right ]\nonumber\\
\tilde{b}-b(u)&=&\frac{1}{u[\tilde{a}-\langle a
\rangle+\Delta(u)]}-\int_u^1\frac{dv}{v^2}
\frac{1}{\tilde{a}-\langle a \rangle+\Delta(v)}
\end{eqnarray}
and
\begin{eqnarray}
\langle a \rangle=\int_0^1a(v)dv&=&s_0\frac{q^2}{k_z^2}-\langle\sigma \rangle\nonumber\\
\tilde{a}-\langle a \rangle+\Delta(u)&=&c(q,k)q^2+z+\Delta(u)\,.
\end{eqnarray}
$B(u)$ can be written, using (\ref {Bab}) and the inversion
formula (\ref {inversion}), as
\begin{equation}\label{Bu}
\half B(u)=\frac{g(u)}{u}-\int_u^1\frac{g(v)}{v^2}
\end{equation}
where
\begin{equation}\label{gu}
g(u)=\sum_{\qq,k}\frac{1}{c(k)q^2+z+\Delta (u)}=
\int\frac{dk}{2\pi}\frac{1}{4\pi c(k)}[\ln
\frac{\Delta_c}{z+\Delta(u)}]+C_1\,.
\end{equation}
and
\begin{equation}\label{C1}
C_1= \int\frac{dk}{8\pi^2c(k)}\ln \frac{c(k)}{c(\pi/d)}\,.
\end{equation}
Here $c(q,k)$ of Eq. (\ref {cqk}) is replaced by $c(k)$ as defined
in (\ref {cqk}, \ref {cc'}) while the $q^2$ term in Eq. (\ref
{cqk}) amounts to redefining the upper cutoff into $q_u^2=4\ln
(a/d)/\lambda_{ab}^2$, (considering $k\approx \pi/d$ as the
dominant range of the following $k$ integration) and
$\Delta_c=c(\pi/d)q_u^2$. In the following a variable $t$ is
temperature in units $T_d^0$ of the pure system (Eq. \ref {Td}),
i.e.
 \begin{equation}\label{t}
 t=\frac{T}{T_d^0}=\int\frac{dk}{16\pi^2}\frac{1}{c(k)}\,.
 \end{equation}
Eq. (\ref {gu}) is then
\begin{equation}\label{gu1}
g(u)=2t\ln \frac{\Delta_c}{z+\Delta (u)}+C_1\,.
\end{equation}

To find $\Delta(u)$ we note that Eq. (\ref {extremumc}) is
equivalent to $\sigma(u)=(E_v/Td)\exp [-B(u)/2]$. Differentiating
this equation and using $\Delta'(u)=u\sigma'(u)$ we obtain
\begin{equation}\label{D'}
\frac{\Delta'(u)}{u}=-\frac{d}{du}[\frac{\Delta'(u)}{g'(u)}]\,,
\end{equation}
which by using (\ref {gu1}) can be written as
\begin{equation}\label{dD}
(\frac{1}{u}-\frac{1}{2t})\frac{d\Delta}{du}=0\,.
\end{equation}
The solution of this equation is a one step function, i.e.
$\Delta(u)$ jumps at $u=2t$ from zero to a constant value
$\Delta_0$ at $2t<u<1$. The solution is therefore nontrivial if
$t<1/2$.

To complete the solution, the function $B(u)$ from (\ref {Bu}) is
needed
\begin{eqnarray}
\half B(u)&=&C_1+\ln \frac{\Delta_c}{z}+(2t-1)\ln
\frac{\Delta_c}{z+\Delta_0}\qquad
u<2t    \nonumber\\
&=& C_1+2t\ln\frac{\Delta_c}{z+\Delta_0} \qquad \qquad  \qquad \qquad
2t<u<1
\end{eqnarray}
which yields for $\sigma(u)$
\begin{eqnarray}\label{sig}
\sigma(u)=\sigma_0&=&\frac{z}{z+\Delta_0}\sigma_1 \qquad \qquad
\qquad u<2t
\nonumber\\
\sigma_1&=&\frac{E_v}{Td}(\frac{z+\Delta_0}{\Delta_c})^{2t}e^{-C_1}
\qquad 2t<u<1   \,.
\end{eqnarray}
Finally, from Eq. (\ref {Du}) we have $z+\Delta_0=2t\sigma_1$, hence,
\begin{equation}\label{D0}
\frac{z+\Delta_0}{\Delta_c}=(\frac{2E_v}{dT_d^0\Delta_c}e^{-C_1})^{\frac{1}{1-2t}}
\end{equation}
A consistent weak coupling solution is indeed possible only at
$t<\half$.

To find a second equation for $z$ from (\ref {extremumb}) we need
the first inversion formula in (\ref {inversion})
\begin{equation}\label{Gtilde}
\tilde{G}(q,k)=\frac{\frac{s_0}{k_z^2c(k)}+\frac{1}{2t}}{c(k)q^2+z}+\frac{\sigma_0-
\frac{s_0z}{k_z^2c(k)}}{(c(k)q^2+z)^2}+\frac{1-\frac{1}{2t}}{c(k)q^2+z+\Delta_0}
\end{equation}
and after the $\qq$ summation
\begin{equation}\label{A1}
A_{\alpha}=\sum_{\qq,k}\tilde{G}(q.k)=\int\frac{dk}{8\pi^2}[\frac{1}{2tc(k)}
\ln \frac{z+\Delta_0}{z}
+\frac{1}{c(k)}\ln \frac{\Delta_c}{z+\Delta_0}
+\frac{\sigma_0}{zc(k)}]
 +C_1+\frac{s_0}{8\pi^2}[I(z)+zI'(z)]\nonumber\\
\end{equation}
where
\begin{equation}\label{Iz}
I(z)=\int\frac{dq^2dk}{k_z^2c(k)}\frac{1}{c(k)q^2+z}
\end{equation}
 and $I'(z)=dI(z)/dz$. For $\Delta_0\neq 0$ we have from Eq. (\ref
{sig}) $\sigma_0=z/2t$
while for $\Delta_0=0$ (possible at $t>\half$ as found below) we
have $\sigma_0\sim z^{2t}\ll z$, hence, with $s$ defined in (\ref {s}),
 \begin{eqnarray}
 A_{\alpha}&=&\ln (2etE_v/z d)+(s_0/8\pi ^2)[I(z)+zI'(z)]  \qquad
\Delta_0\neq 0 \label{A2}\\
 &=&C_1+2(t+s)\ln \frac{\Delta_c}{z} +2s  \qquad \qquad \qquad \qquad
\Delta_0 = 0\,. \label{A3}
 \end{eqnarray}

Formally $I(z)$ diverges at $k=0$; this divergence can be traced
back to our assumption that the $\cos [{\bf Q}\cdot ({\bf
u}_{l}^{n,\alpha} -{\bf u}_{l}^{n,\beta})]$ term is expanded into
the $s_0$ term in Eq. (\ref{Hreplica}). Retaining this cosine
leads to domains of correlated ${\bf u}_l^n$. In Appendix A the
joint BG-JG solution is found and is shown to remove the
$k\rightarrow 0$ divergence. A combined BG with non-singular phase
solution is also shown in Appendix B to remove this divergence.
The presence of BG, however, produces a term $\sim \ln^2z$ in a
regime near decoupling (Eq. \ref{ln2} in the regime of Eq.
\ref{cond2}). This "anharmonic" regime is studied further in
section IV. Excluding this anharmonic regime, the dominant part of
$I(z)$ is
\begin{equation}\label{Iz3}
I_0(z)=2\int_{1/a}^{\pi/d}\frac{dk}{k_z^2c(k)}\int
\frac{dq^2}{c(k)q^2+z}=\frac{\pi
d}{4c^2(\pi/d)}\ln\frac{\Delta_c}{z}
\end{equation}
The $I(z)$ term in Eq. (\ref {A1}) can then be written as
\begin {equation}\label{Iz1}
\frac{s_0}{8\pi ^2}I(z)\approx \frac{s_0}{8\pi ^2}I_0(z)=2s\ln
\frac{\Delta _c}{z} \,.
\end {equation}
Therefore, the renormalized Josephson coupling of
Eq.~(\ref{extremumb}) is for $\Delta_0\neq 0$, using Eq. (\ref
{A2}),
\begin{equation}
\frac{z}{\Delta _c}=e^{-1}(\frac{E_J^2}{2T t
dE_v\Delta_c})^{\frac{1}{1-2s}} \,. \label{z}
\end{equation}
Note that $E_v$ is generated from $E_J$ by RG \cite{H3,Scheidl},
i.e. $E_v\sim E_J^2$ initially; however, $E_v$ is RG relevant at
$t<\half$ even in 1st order RG (Eq. \ref {RG}), hence we consider
$E_v$ and $E_J$ as comparable so that $E_J^2/(2T t
dE_v\Delta_c)\ll 1$. Hence a consistent weak coupling
$z/\Delta_c\ll 1$ solution is possible only for $s<\half$. Thus
$s=\half$ marks a disorder induced decoupling with $z=0$ at
$s>\half$.

Comparing Eqs.~(\ref {D0},\ref{z}) shows that $\Delta_0$ vanishes
at $s=t$ (up to $O[\ln (E_J/E_v)/\ln E_v]$ term, small for
$E_J\approx E_v\ll d\Delta_cT_d$). Formally there is a solution
with $\Delta_0<0$ when $s<t$. However, the average distribution
\cite{Mezard} of $|b({\bf q},k)|^2$, which is $\sim \exp [-|b({\bf
q},k)|^2/G_{\alpha,\alpha}({\bf q},k)]$, is acceptable only if
$G_{\alpha,\alpha}({\bf q},k)>0$; this is therefore a
thermodynamic stability criterion. Note in particular from Eq.
(\ref {Gtilde})
\begin{equation}
G_{\alpha\alpha}(q=0,k)=\frac{1}{2t}(\frac{2}{z}-\frac{1-2t}{z+\Delta_0})\,.
\end{equation}
When $s<t$ and $\Delta_0<0$ the power dependence in Eq. (\ref
{D0}) implies that $z+\Delta_0 \ll z$ (unless too close to $t=s$,
i.e. $s-t\sim 1/|\ln E_v|$) and therefore
$G_{\alpha\alpha}(q=0,k)<0$. This shows that only $\Delta_0>0$ is
acceptable.

Thus the regime where both $z$, $\Delta_0$ are finite is limited
to $s<\half$, $t<s$; we term this regime the coupled Josephson
Glass (JG) phase. The "coupled" notation means that the
renormalized Josephson coupling is finite, i.e. $z\neq 0$. The
glass parameter vanishes (continuously) at $t=s$ while the
Josephson coupling vanishes (with an apparent discontinuity - see
section IV ) at $s=\half$ (see Fig. 1).
 For $s>\half$ and $t<\half$ the solution is $z=0$
while $\Delta_0 \neq 0$ satisfies Eq.~(\ref{D0}), i.e. it is a
{\em decoupled} JG phase. Recall that the JG order parameter
$\Delta_0$ is due to $E_v$ which is initially generated by $E_J$.
In fact, the RG of Eq. (\ref{RG}) shows (see a similar effect in
Fig. 3 of the companion article \cite{GH} for the pure system)
that $E_J$ first increases (scaling from $\xi_0$ to $1/q_u$),
generating the $E_v$ term, and only at scales beyond $1/q_u$ $E_J$
decreases to zero. It is remarkable then that $E_J$ is
renormalized to zero while the JG order survives, much like the
smile of the Cheshire cat.

\begin{figure}[b]
\begin{center}
\includegraphics[scale=0.7]{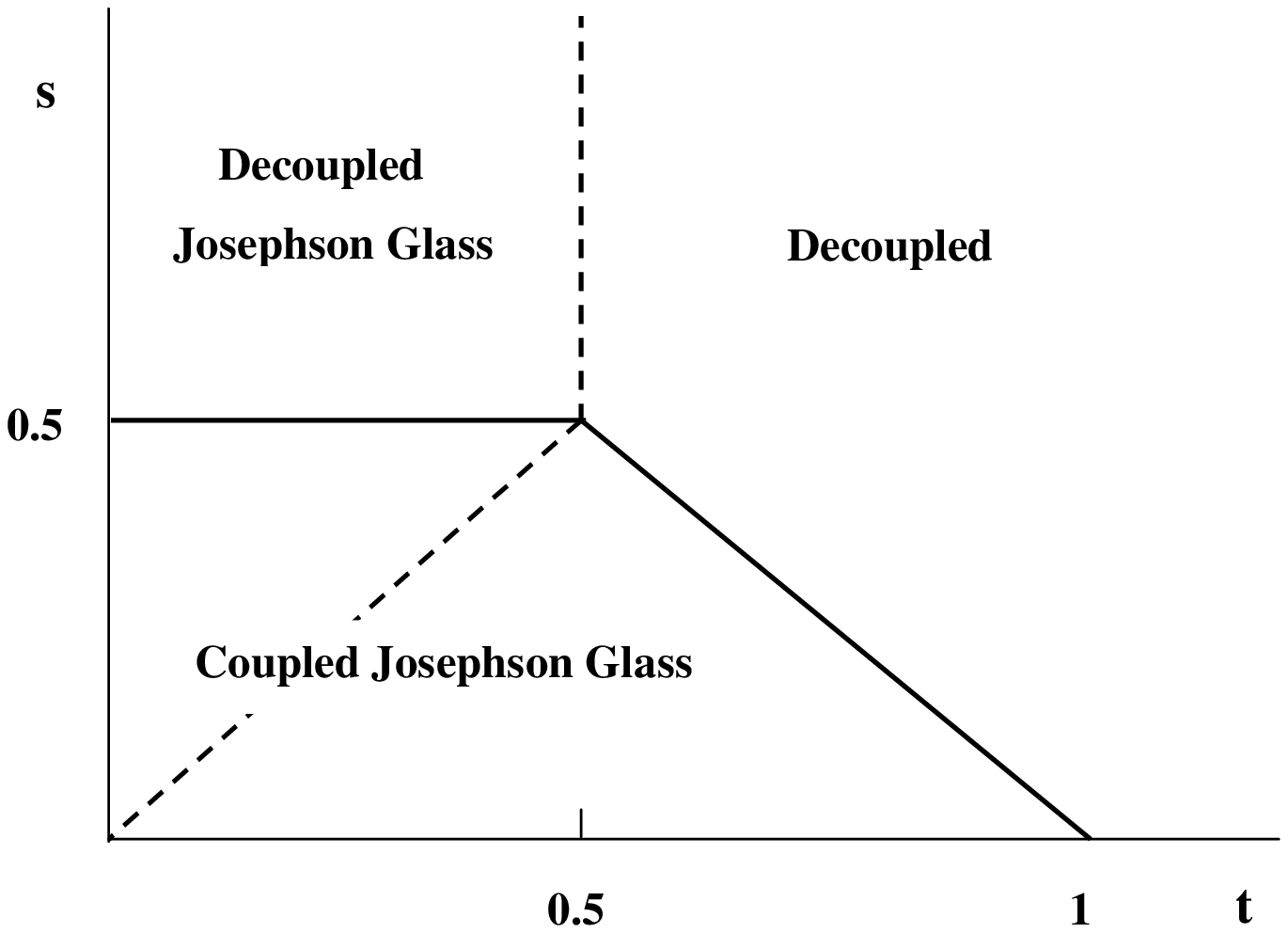}
\end{center}
\caption{Phase diagram. Full lines are decoupling lines where the
Josephson coupling vanishes. The upper dashed line is a depinning
transition where the Josephson glass parameter vanishes; the lower
dashed line is either a 1st order line or a crossover into a weaker JG phase, i.e. weaker
pinning.\vspace{1.5cm}}
\end{figure}

Finally, for $\Delta_0=0$ a replica symmetric solution is valid at
$s<t$, which upon using Eqs. (\ref {extremumb}, \ref{A3}) becomes
\begin {equation}\label{z1}
\frac{z}{\Delta_c}=(\frac{E_J}{Td\Delta_c}e^{-s-\half
C_1})^{\frac{1}{1-s-t}} .
\end{equation}
Thus $s+t=1$  for $s<\half$ defines a "thermal" decoupling
transition.

The interpretation of the phase diagram needs to be supplemented
by a few observations from an RG analysis. The RSB results above
coincide with those of a 2D model where the parameters $t,s$ of
the 3d system, as suitable $k$ averages (Eqs. \ref {t},\ref
{Iz1}), correspond to Hamiltonian parameters of the 2D system
\cite{H3}. With this correspondence in mind, we infer next some
qualitative modifications by using the 2D RG equations
\cite{H3,Scheidl}, Eq. (\ref {RG}). Note first that in a coupled
phase $z$ is RG relevant and therefore $E_v$, which is generated
to order $z^2$, is finite too,  hence a weak glass phase is
expected also in the regime $s<t<1-s$; this weak glass order is
not captured by the RSB solution. The line $t=s$ for $s<\half$ can
therefore be either a 1st order transition or a crossover
line. RG suggests (Eq. \ref{RG}) this crossover line at $t=\half$: at $t<\half$
RG yields $E_v$ which is largely independent of $z$, hence a
strong JG order, while at $t>\half$ RG generates $E_v\sim z^2$
with a weak JG order. The stability of the RSB solution shows that
in fact this line, which is either 1st order or a crossover, is at $t=s$.

The RG, shows also a disorder induced decoupling, since Eq. (\ref
{RG}) has a fixed point with $u^*=0$ and $v^*=(1-2t)/\gamma' t$,
stable at $t<\half$ and strong disorder. Note that for this
solution $s\sim \ln \xi$ increases with scale $\xi$, hence the
correlator $\Gamma (r)=\langle \cos b_n^{\alpha}(\rr) \cos
b_n^{\alpha}(0)\rangle$ which by RSB decays as $r^{-2-4s}$ is
actually decaying faster as $\ln \Gamma (r) \sim -\ln^2r$.
Explicit solution of the 2D RG equations \cite{Scheidl} found
indeed a phase diagram very similar to that in Ref.
\onlinecite{H3} or in Fig. 1.

\begin{figure}[htb]
\includegraphics[scale=0.7]{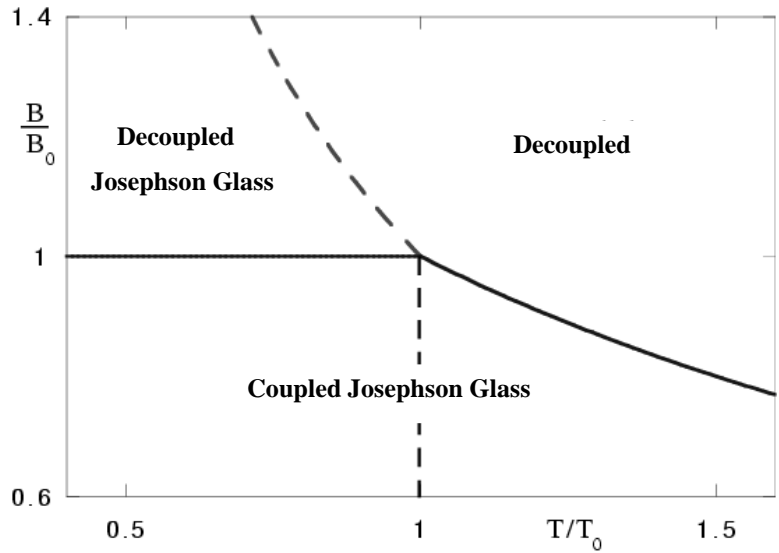}
\caption{Phase diagram in terms of field and temperature. Full
lines are decoupling lines [$B=B_0$ and $B=2B_0T_0/(T+T_0)$] where
the Josephson coupling vanishes. The upper dashed line is a
depinning transition ($T=B_0T_0/B$) where the Josephson glass
parameter vanishes; the lower dashed line ($T=T_0$) is either a 1st order transition or a crossover
into weaker pinning.\vspace{1.5cm}}
\end{figure}

The phase diagram, shown in Fig. 1, has three phase transition
lines and a line which is either a 1st order or a crossover line. All these lines meet at a multicritical
point $s=t=1/2$. We interpret the transition where $\Delta_0$
vanishes as a depinning transition, i.e. the JG order parameter
provides an additional pinning to that from the Bragg glass. The
phase diagram has then a decoupling line which crosses a depinning
line at the multicritical point. The decoupling line has a
disorder driven section, $s=\half$ at $t<\half$.

The phase diagram in terms of field and temperature is derived by
defining $B_0, T_0$ as the field and temperature value of the
multicritical point and is shown in Fig. 2. $B_0$ is determined by
the disorder strength via $s=\half$ while $T_0=\half T_d^0
(a=\sqrt{B_0/\phi_0})$ (Eq. \ref{Td}). Hence $s=B/2B_0$ and
$t=TB/T_0B_0$, up to $\ln B$ terms. Since s increases with $B$ the
$s=\half$ line defines a decoupling transition from a coupled JG
phase at low $B$ to a decoupled JG phase at high fields.

The coupled JG phase at $B<B_0$ goes through either a 1st order or a crossover line
at $t=s$, i.e. at $T=T_0$ (up to $\ln B$ factors). Therefore at
$T>T_0$ the glass parameter $\Delta_0$ is significantly reduced
implying  depinning,  a change from strong to weak pinning.
 The decoupled JG phase undergoes a depinning transition into a
decoupled phase at $T=B_0T_0/B$. Note that all phases, even the
high $T$ decoupled one, are Bragg glass phases of the flux
lattice; in the decoupled phase the lattice is maintained by the
interlayer electromagnetic coupling.

The JG coupled phase at $T>T_0$ undergoes a decoupling transition at
$t=1-s$, i.e.
 $B=2B_0T_0/(T+T_0)$. This transition is continuous;
the variational method of the pure system has been formally
extended to higher $J/T$ and found to be of first order
\cite{Daemen}. As shown in the companion article \cite{GH}, the transition
remains 2nd order when proper 2nd order RG is employed. Disorder,
however, leads to an apparent discontinuity near decoupling, as
discussed in the next section.

\section{Domain sizes}
In this section we estimate various domain sizes and evaluate
displacement fluctuations which identify these sizes. Remarkably,
the expressions for the domain sizes are confirmed (up to
numerical prefactors) by BG solutions (Appendices A-C). The
nonsingular phase, which was irrelevant for the purpose of the
phase diagram in section III, is essential now.

To appreciate the effect of the nonsingular phase $\theta$, we
briefly review the derivation of the transverse tilt modulus
$c_{44}$ of a pure flux lattice \cite{GH2}. The Josephson phase
involves the contribution of pancake fluctuations via $b_n(\rr)$
as well as a nonsingular phase, with the Hamiltonian Eq. (\ref
{Hpure2}). To identify $c_{44}$ we expand the Josephson coupling
to 2nd order in  \mbox{${\tilde
b}(\qq,k)=b(\qq,k)+\theta(\qq,k)$},
\begin {eqnarray}\label{Hpure3}
{\cal H}_{pure}/T =&& \half\sum_{q,k} \{[G_{f}^{-1}(q,k)
+\frac{E_J}{Td}]\,\left |\, {\tilde b}(\qq,k)
-b(\qq,k)\frac{G_{f}^{-1}(q,k)}{G_{f}^{-1}(q,k)+
\frac{E_J}{Td}}\,\right |^2+ \nonumber\\
&&\frac{G_{f}^{-1}(q,k) \frac{E_J}{Td}}{G_{f}^{-1}({\bf
q},k)+\frac{E_J}{Td}}|b(\qq,k)|^2 +\half c(q,k)|b(\qq,k)|^2
+O({\tilde b}^4(\qq,k))\}\,.
\end{eqnarray}
The first term decouples from $b(\qq,k)$ and with $|b(\qq,k)|^2=
(2\pi d^2)^2k_z^2|u^{tr}(q,k)|^2/q^2$  (Eq. \ref {b}) we identify
\cite{GK,Sudbo,GH2}
\begin{equation}
c_{44}(q,k) = c_{44}^0(k)+ \frac{B^2 }{4\pi}
\frac{1}{1+\lambda_c^2 q^2 + \lambda_{ab}^2 k^2}
+\frac{2B\phi_0}{(8\pi
\lambda_c)^2} \ln (a^2/4\pi \xi_0 ^2)  \label{c44}
\end{equation}
where $\lambda_c^2=\lambda_{ab}^2\tau/(4\pi d^2E_J)$; the last term
is from
reducing high momenta of the 2nd term of (\ref {Hpure3}) into the 1st
Brilluin zone.

 The second term of $c_{44}(q,k)$ is peculiar: at
$q\neq 0$ it vanishes when $E_J$ vanishes and $\lambda_c
\rightarrow \infty$, as it should. However, at $q=0$ this term
seems to survive even if $\lambda_c \rightarrow \infty$. The
origin of this peculiarity is that the harmonic expansion of the
Josephson cosine term which identifies $c_{44}$ fails\cite{GH2}
when both $q, \,1/\lambda_c \rightarrow 0$. The shift in the 1st
term of Eq. (\ref {Hpure3}) identifies an expansion parameter
\cite{GH2} with terms $\sim q^2k_z^2|u_{T}({\bf q},k)|^2/
[q^2+\lambda_c^{-2}(1+\lambda_{ab}^2k_z^2)]^2$, which diverge when
both $q,1/\lambda_c \rightarrow 0$ and the expansion becomes
invalid. In fact, the nonlinear cosine term replaces $E_J/Td$ by
$z$ or $\lambda_c$ is replaced by a renormalized
\begin{equation}\label{lamcR}
\lambda_c^R=\sqrt{\lambda_{ab}^2\tau/(4\pi Td^3z)}
\end{equation}
which diverges at decoupling. Hence usual elasticity at
$q,1/\lambda_c^R\rightarrow 0$ near decoupling is ill defined.

 The Bragg glass domain size $R_{BG}$ (parallel to the layers) sets a
scale for the relevant $q$ values. When $R_{BG} > \lambda_c^R$ the
tilt modulus is large, containing the $B^2/4\pi$ term of Eq.\
(\ref{c44}). However, as decoupling at the field $B_0$ is
approached $\lambda_c^R$ diverges so that when $R_{BG} <
\lambda_c^R \,$ Eq.\ (\ref{c44}) fails to describe $c_{44}$ on the
scale of $q\approx 1/R_{BG}$. This defines an anharmonic crossover
regime where usual elasticity cannot be used to derive Bragg glass
properties. Finally, at $B>B_0$ elasticity is restored and
$c_{44}$ is reduced to the first term in Eq.\ (\ref{c44}). The
main interest is in the regime of strong fields, i.e. $a\lesssim
2\lambda_{ab}$ where $T_0\ll \tau$ is below melting. Thus at
$B<B_0$ and for sufficiently large domains the second term in Eq.
(\ref {c44}) dominates and $c_{44}=B^2/4\pi$ while at $B>B_0$ only the magnetic coupling survives $c_{44}=c_{44}^0(k)$ which at $ka\ll 1$ becomes $\tau/(32\pi \lambda_{ab}^2 d)$. Hence there is an apparent discontinuity,
\begin{eqnarray}
c_{44} &=& \pi \lambda_{ab}^2\tau/da^4    \hspace{18mm}
\lambda_c^R <
R_{BG} \label{c44a}\\
    &=& \tau/(32\pi \lambda_{ab}^2 d)   \hspace{15mm}
\lambda_c^R = \infty  \label{c44b}
\end{eqnarray}
Thus $c_{44}$ is reduced within the anharmonic regime by the
small factor
\begin{equation}\label{epsilon}
\epsilon=a^4/(32\pi^2\lambda_{ab}^4)\,.
\end{equation}

The apparent discontinuity in $c_{44}$ affects also the domain
sizes which can be estimated by a dimensional argument
\cite{LO,Giamarchi}. Consider the tilt $c_{44}$ and shear $c_{66}$
terms of the elasticity Hamiltonian for the displacement ${\bf
u}({\bf r})$ and its transverse component ${\bf u}_T({\bf r})$.
Rescaling parallel and perpendicular lengths yields an isotropic
form \cite{Blatter,Natterman}, which together with the pinning
energy (\ref {Hpin2}) yield (ignoring elasticity of longitudinal
displacements)
\begin{equation}
{\cal H}= \int d^3r\{\half c_{44}^{1/3}c_{66}^{2/3} [
{\mbox{\boldmath $\nabla$}}u_T({\bf r})]^2
-(\xi_0^2/a^2d)U_{pin}({\bf r}) \sum_{{\bf Q}}\cos {\bf Q}\cdot
[\rr- {\bf u} ({\bf r})]\} \label{H3}
\end{equation}
 where the disorder coupling to
${\mbox{\boldmath $\nabla$}}u_T({\bf r})$ is neglected. To
estimate the energy gain from disorder we consider the overlap of
the disorder energy between two configurations ${\bf u} ({\bf r})$
and ${\bf u}' ({\bf r})$ which are solutions for two realizations
of the random potential \cite{Blatter}; this overlap is a measure
of the energy variance in configuration space. The ${\bf r}$
integration leads to a single ${\bf Q}$ sum so that the variance
is \mbox{$\sim \sum_{{\bf Q}}\cos {\bf Q}\cdot[{\bf u} ({\bf r})-
{\bf u}' ({\bf r})]$}. Each of  ${\bf u} ({\bf r})$ and ${\bf u}'
({\bf r})$ has fluctuations $\langle u^2 \rangle \approx \langle
u_T^2 \rangle$ in a domain of size $R'$ so that the ${\bf Q}$ sum
is cutoff by $Q \lesssim \langle u_T^2 \rangle ^{-1/2}$. Below
this cutoff the cosine can be expanded and summed so that
averaging Eq.\ (\ref{H3}) yields
\begin{equation}\label{HR'}
\langle H \rangle /R'^3 = \half c_{44}^{1/3}c_{66}^{2/3}
\langle u_T^2 \rangle R'^{-2} - {\bar U}^{1/2}\xi_0^2
/[a^2d\langle u_T^2 \rangle R'^3]^{1/2} \, .    \label{Hav}
\end{equation}
Minimizing with respect to $R'$ yields $R' \sim \langle u_T^2
\rangle ^3$, i.e. the Flory exponent \cite{Giamarchi}. This exponent is not exact; the more
accurate statement, shown within the BG solution \cite{Giamarchi},
is that the disorder averaged correlation ${\tilde B}(R')=\langle
[u_T({\bf R}')-u_T(0)]^2\rangle\sim R'^{1/3}$ is a quantitatively
correct description in the range between the pinning length $R_p$
where ${\tilde B}(R_p)=\xi_0^2$ and $R_{BG}$ where ${\tilde
B}(R_{BG})=a^2$. The fluctuations $\langle u_T^2 \rangle$ on scale
$R'$ in the dimensional argument correspond then to ${\tilde
B}(R')$.

\begin{figure}[htb]
\begin{center}
\includegraphics[scale=0.7]{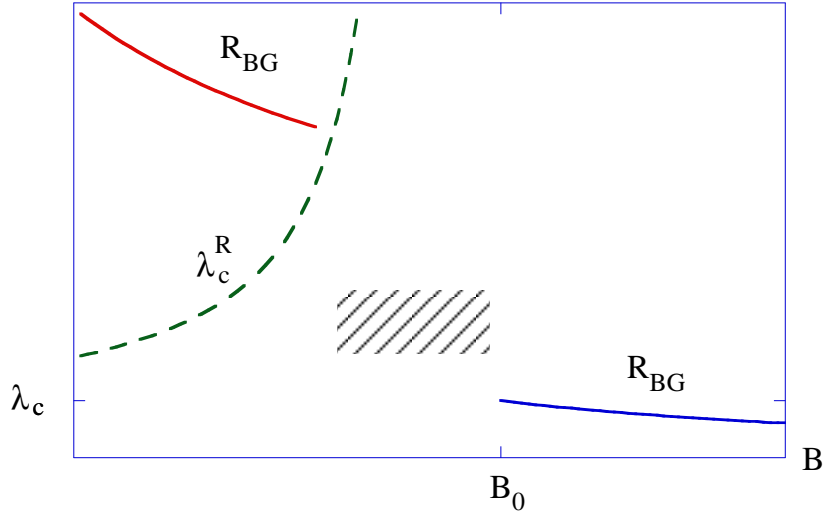}
\end{center}
\caption{ Bragg glass domain size $R_{BG}$ parallel to the layers
and the renormalized London length perpendicular to the layers
$\lambda_c^R$; the latter diverges at the decoupling field $B_0$.
$R_{BG}$ can be found from elasticity for $B<B_0$ only if
$R_{BG}>\lambda_c^R$; otherwise, as in the hatched region, the
elastic tilt modulus is ill defined. \vspace{1.5cm}}
\end{figure}

The domain size parallel to the layers is, from minimizing Eq.
(\ref{HR'}), (up to $\ln(a/d)$ and a numerical prefactor)
\begin{eqnarray}\label{R}
R^+ & \approx & (\lambda_{ab}/a)^5 \langle u_T^2 \rangle ^3
/(s\xi_0^4 d)
 \hspace{20mm} \lambda_c^R < R^+
\nonumber\\
R^-  & \approx & (\lambda_{ab}/a)^3\langle u_T^2 \rangle ^3 /(4\pi
s\xi_0^4d) \hspace{15mm}  \lambda_c^R = \infty \,.
\end{eqnarray}
The pinning length $R=R_p$ is given by Eq.\ (\ref{R}) with
$\langle u_T^2 \rangle \approx \xi_0^2$. The condition $\lambda_c^R < R_p^+$ is not valid for BSCCO parameters; to allow for large
pinning domains one needs either $a\ll \lambda_{ab}$ or to allow
for domains with a somewhat larger fluctuations in $\langle u_T^2
\rangle$; the latter increases $R_p$ very rapidly  since it
increases with the 6-th power of $u_T$. The critical current can
now be estimated \cite{Blatter,LO} by balancing the Lorenz force
$j_cBR^3/c$ with the pinning force $\langle H \rangle /\xi_0$
(evaluated at the minimum of Eq.\ (\ref{Hav})), leading to
$j_c\sim 1/c_{44}$. Increasing the field within the anharmonic
regime decreases $c_{44}$ by the factor $\epsilon$ so that $j_c$
is enhanced by a $1/\epsilon$ factor which is significant when
$a\lesssim \lambda_{ab}$.

A second length scale $R=R_{BG}$ is identified by  Eq. (\ref{R})
with the fluctuations $\langle u_T^2 \rangle \approx a^2$.
 The proper definition of $R_{BG}$ is the scale for the onset of the
 $\ln r$ form for the displacement correlation function, as
 inferred in Eq. (\ref {RBG}) or (\ref {R+-}). It is remarkable that
Eq. (\ref
 {R}) gives the correct form for for $R_{BG}$, up to a numerical
 prefactor, i.e. Eqs. (\ref {RBG}, \ref {R+-}). Eq. (\ref {R})
 shows that $R_{BG}$ is reduced by $\epsilon ^{1/2}$ through the
anharmonic regime. The onset of the anharmonic regime is at
$R_{BG}^+\approx \lambda_c^R$, i.e.
\begin{equation}\label{cond1}
\lambda_c^R\approx 10^{-3}\frac{a\lam^5}{sd\xi_0^4}
\end{equation}
with a numerical prefactor from the BG solution (Eqs. \ref {RBG},
\ref {R+-}). For BSCCO or YBCO parameters at $s\approx \half$ this
reduces to $\lambda_c^R/\lam\approx 10^5$, i.e. the initial
anisotropy of $\lambda_c/\lam =10-100$ has to increase to $\approx
10^5$. Since $z$ is exponentially renormalized (Eqs. \ref {z},
\ref {z1}) this anharmonic range may be observable.

Fig. 3 illustrates the lengths $R_{BG}$ and $\lambda_c^R$,
demonstrating the anharmonic regime within which $R_{BG}$ has a
significant drop and correspondingly $j_c$ has an apparent jump.
Note that even in the decoupled phase ($B>B_0$) $R_{BG}$ is large
for typical type II superconductors,
 $R_{BG}\approx \lambda_{ab}^3 a^3 /(4\pi s \xi_0^4 d) \gg a$,
consistent with a decoupling transition within the Bragg glass phase, i.e. below
a melting transition.

The solution of section III can also be extended to include the
nonsingular phase. Since disorder is linearized, the pinning
length $R_p$ can be determined, though the BG length cannot.
Appendix C develops this solution and shows that in
the coupled phase $R_p$ coincides with $R^+$ Eq. (\ref {R}) (with
$\langle u_T^2 \rangle \approx \xi_0^2$), up to a numerical
prefactor.

The main result is then that the fluctuations in $u_T(r)$ behave
with an effective $c_{44}$ which is large when $q<1/\lambda_c^R$
(Eq.\ (\ref{c44a})), i.e. for domain sizes $R_{BG}>\lambda_c^R$,
while for $z=0$ $c_{44}$ is reduced (Eq.\ (\ref{c44b})). While the condition $\lambda_c^R < R_p^+$  in Eq.  (\ref{R}) is not valid for BSCCO (the pinning domains are likely to be two dimensional) our results for the anharmonic regime itself in terms of the much larger $R_{BG}$ are valid. The existence of a narrow anharmonic regime leads to an apparent jump in $c_{44}$ which possibly affects the critical current.

In the anharmonic region below decoupling (see Fig. 3) where
$R_{BG}<\lambda_c^R$ a more complete form [e.g. Eq.\ (\ref{fluc})]
is required to interpolate between the limiting forms of $c_{44}$. However, a method
relying on an effective harmonic theory, such as RSB, is suspect
within the anharmonic regime, since the system has no effective
elastic constants. Furthermore, RSB signals this deficiency by
producing a $ln^2z$ term, precisely in the the anharmonic regime
found here, as shown in Appendices A and B.

\section{Josephson Plasma resonance}

Josephson plasma resonance provides extremely useful data for identifying
phases of vortex matter
\cite{Matsuda,Koshelev,Shibauchi,Matsuda2}. In particular a jump
in the resonance frequency $\omega_{pl}$ has shown
\cite{Shibauchi,Matsuda2} that the Josephson coupling is strongly
modified at the second peak transition. In this section we derive
$\omega_{pl}$ in the ordered phases and also consider the
fluctuation contribution in the disordered phase. The Josephson plasma frequency is given by \cite{Koshelev} (see also the companion article \cite{GH} section V)
\begin{equation}\label{opl}
\omega_{pl}^2=\frac{16\pi e^2dE_J}{\epsilon_0 \hbar^2}\langle \cos b
\rangle
\end{equation}
where $\epsilon_0$ is a dielectric constant. The task is then to evaluate the thermodynamic 
average $\langle \cos b \rangle$.

Consider first the ordered phases where at least one of $z$ and
$\Delta_0$ is finite. We start by evaluating ${\cal F}_{var}$ of
Eq. (\ref {Fvar}) for a general one step RSB, recover the solution
of section III, and then identify $\langle \cos b \rangle$. This
derivation is needed so that the free energy itself can be
evaluated, and from the latter $\langle \cos b \rangle$ is
inferred. The self mass term $\sigma_{\alpha,\beta}$ of Eq. (\ref
{extremuma}) is written for a one step solution in the form
\begin{equation}
{\hat \sigma}=\sigma_0{\hat L}+(\sigma_1-\sigma_0){\hat
C}-[\sigma_0n+(\sigma_1-\sigma_0)m]{\hat I}
\end{equation}
where ${\hat L}_{\alpha,\beta}=1$ and ${\hat C}$ has $1$ elements
in blocks of size $m\times m$ sitting consecutively along the
diagonal, and $0$ elements otherwise. For $n\rightarrow 0$ we
identify $(\sigma_1-\sigma_0)m=\Delta_0$ so that
\begin{equation}
{\hat G}(q,k)=[(c(q,k)q^2+z+\Delta_0){\hat
I}+(-s_0\frac{q^2}{k_z^2}-\sigma_0){\hat
L}-\frac{\Delta_0}{m}{\hat C}]^{-1}\equiv \alpha {\hat I}+\beta
{\hat L}+\gamma {\hat C}\,.
\end{equation}
It is straightforward to identify the coefficients of the inverse
matrix
\begin{eqnarray}
\alpha(q,k)&=&\frac{1}{c(q,k)q^2+z+\Delta_0}\,, \qquad \qquad
\alpha=\sum_{\qq,k}\alpha(q,k)=2t\ln
\frac{\Delta_c}{z+\Delta_0}+C_1\nonumber\\
\beta(q,k)&=&\frac{s_0\frac{q^2}{k_z^2}+\sigma_0}{[c(q,k)q^2+z]^2}\,,
\qquad \qquad \beta=\sum_{\qq,k}\beta(q,k)=2s(\ln
\frac{\Delta_c}{z}-1)+\frac{2t\sigma_0}{z}\nonumber\\
\gamma(q,k)&=&-\frac{1}{m}[\alpha(q,k)-\frac{1}{c(q,k)q^2+z}]
\,,\qquad \gamma
=\sum_{\qq,k}\gamma(q,k)=\frac{2t}{m}\ln\frac{z+\Delta_0}{z}
\end{eqnarray}
where the form of Eq. (\ref {Iz1}) is used for $I(z)$ in the 2nd
line. The definition of ${\hat \sigma}$ identifies
$\sigma_1=2E_ve^{-\alpha}/d$ and
$\sigma_0=2E_ve^{-\alpha-\gamma}/d$. We follow a similar algebra
in section IV of Ref. \onlinecite{H3} to evaluate the free energy
density per replica as
\begin{equation}\label{frep}
f(m,z,\Delta_0)=f_0+(1-\frac{1}{m}t\Delta_0+(1+\frac{s}{t})tz-
\frac{E_v}{Td}(2t-m)e^{-\alpha-\gamma}+\frac{E_v}{Td}(1-m)e^{\alpha}-
\frac{E_J}{Td}e^{-\half(\alpha+\beta+\gamma)}
\end{equation}
where $f_0$ is $m$, $z$ and $\Delta_0$ independent. Minimizing
$f(m,z,\Delta_0)$ yields $m=2t$ and Eqs. (\ref {sig},\ref
{D0},\ref {z})for $\sigma_0,\sigma_1,z+\Delta_0$ and $z$. A
replica symmetric solution is also possible with $\Delta_0=0$
leading to Eq. (\ref {z1}). The free energy at minimum is
\begin{equation}\label{fmin}
f_{min}=f_0+(t-1+\frac{1}{4t})(\Delta_0+z)+(s-\half)z\,.
\end{equation}

The Hamiltonian Eq. (\ref {Hreplica}) shows that $\langle \cos b
\rangle=-Td(\partial f/\partial E_J)$. As discussed below Eq.
(\ref {z1}) $E_v$ is generated from $E_J$ in 2nd order RG  so that
$E_v\sim E_J^2$ initially, while $E_v$ is RG relevant at
$t<\half$, so that its value which is to be used by the
variational scheme is more weakly $E_J$ dependent. We assume then
$E_v\sim E_J^{\kappa}$ with $0<\kappa<2$. Hence in the
$\Delta_0\neq 0$ phases
\begin{eqnarray}\label{cosb1}
\frac{\partial (z+\Delta_0)}{\partial E_J}&=&\frac{\kappa
(z+\Delta_0)}{E_J(1-2t)}\nonumber\\
\frac{\partial z}{\partial
E_J}&=&\frac{(2-\kappa)}{(1-2s)E_J}\nonumber\\
\langle \cos
b\rangle&=&-\kappa(1-2t)\frac{z+\Delta_0}{z_{bare}}+(1-\half
\kappa)\frac{z}{z_{bare}}
\end{eqnarray}
where $z_{bare}=E_J/Td$ is the bare value of $z$. For the
$\Delta_0=0$ phase
\begin{eqnarray}\label{cosb2}
\frac{\partial z}{\partial E_J}&=&\frac{z}{(1-t-s)E_J}
\nonumber\\
\langle \cos b \rangle&=&\frac{z}{z_{bare}}
\end{eqnarray}
so that at $T=s=0$ the order is maximal, $\langle \cos b
\rangle=1$.

These results show that the JG order produces a negative
contribution to $\langle \cos b \rangle$ so that when crossing a
depinning line $\langle \cos b \rangle$ is enhanced by the $\sim
\kappa$ terms in Eq. (\ref {cosb1}). Since $\Delta_0$ is
continuous, the jump at depinning is $\kappa(\frac{3}{2}-2t)
z/z_{bare}$. As discusses in section III, the depinning in the
lower part of Fig. 1 is not a strict phase transition, but rather
a crossover line, hence we expect a smeared jump of $\langle \cos
b \rangle$. An observation of a $\langle \cos b \rangle$
enhancement when crossing the lower depinning line at $T\approx
T_0$ ($B<B_0$) would be a clear signature that depinning relates
to JG order. The actual enhancement depends on $\kappa$, for which
we do not have a precise derivation.

Near the decoupling transitions, the presence of anharmonic
regimes, shown in section IV, lead to an apparent jump in $\langle
\cos b \rangle$. This jump relates to the $z$ terms in (\ref
{cosb1}) and also depends on the fluctuation contribution which is
considered next.

We proceed to evaluate fluctuation contribution when $\langle \cos
b \rangle$ is small. As shown by Koshelev \cite{Koshelev} the
local $\langle \cos b_n(r)\rangle$ is finite even at high
temperatures, e.g. above the decoupling transition. The high temperature expansion, while formally ill defined, does reproduce the RG results for $\langle \cos
b \rangle$, as shown in section III of the companion article \cite{GH}. The high
temperature expansion yields 
\begin{eqnarray}\label{cosb3}
\langle \cos b_n(r)\rangle &=& (E_J/2T)\int d^2r\exp
[-A(r)]\nonumber\\
A(r)&=&\sum_{{\bf q},k}(1-\cos {\bf q}\cdot {\bf r})\langle
|b^{\alpha}({\bf q},k)|^2\rangle\,.
\end{eqnarray}
For $r^2>1/q_u^2$ we can use the form (\ref {A1}) with $z$
replaced by a cutoff $c(k)/r^2$ while for $r<1/q_u$ we expand
$1-\cos \qq\cdot \rr \rightarrow \frac{1}{4}q^2r^2$, hence
\begin{eqnarray}\label{AA}
A(r)&=&4(s+t)\ln (q_ur)  \qquad   r>1/q_u\nonumber\\
    &=&\half (s+t)q_u^2r^2  \qquad  r<1/q_u \,.
\end{eqnarray}

The two regimes in Eq. (\ref{AA}) give comparable results, though the $r>1/q_u$ is larger near the transition and reproduces the form of the RG result, as discussed in section III of the companion article \cite{GH}. The latter yields, in terms of the multicritical point coordinates (up to $\ln B$ terms),
\begin{equation}\label{cosb4}
\langle \cos b\rangle \approx\frac{\pi E_J\lambda_{ab}^2} {8\ln
(a/d)}\cdot \frac{B_0T_0}{T(BT_0+BT-B_0T_0)}  \,.
\end{equation}
Well above decoupling at $s+t\gg \frac{1}{2}$ we obtain $\langle \cos b\rangle \sim [BT(T+T_0)]^{-1}$.
A $1/BT$ dependence has been obtained by Koshelev \cite{Koshelev}
with a weakly temperature dependent prefactor for an XY model,
i.e. infinite $\lambda_{ab}$ model. This result corresponds, in fact, to the melted, or liquid phase \cite{GH}. Data on BSCCO \cite{Matsuda}
has shown that $\langle \cos \tilde{b}_n(r)\rangle \sim
B^{-0.8}T^{-1}$ in reasonable agreement with the $1/BT$ form. The
present result shows that in the decoupled phase, below melting, $\omega_{pl}\sim [BT(T+T_0)]^{-1}$, or the form (\ref{cosb4}) near decoupling. This distinct temperature dependence can be used to identify the decoupled phase.

As decoupling is crossed, we
expect a positive fluctuation term to compensate the negative
contribution of the JG order. Thus the forms (\ref
{cosb1},\ref{cosb2}) can be used for the jumps of $\langle \cos
b\rangle$ across depinning or decoupling, while (\ref {cosb3}) is
valid in the high temperature or high field regime where $\langle
\cos b\rangle$ is small.

\section{Discussion}

The present work exhibits the JG order parameter as well as the
decoupling transition with disorder. We discuss now our proposal
for each of the 4 transition lines emanating from the
multi-critical point (Fig. 2) and compare with experimental data.

Consider first the decoupling transition within the JG phase at
$B=B_0$, $T<T_0$. We have shown that RSB methods are suspect
within a narrow region near decoupling, where usual elasticity is
ill defined (Fig. 3). RSB identifies this as a $\ln^2z$ divergence
in $A_{\alpha}$ which renormalizes $z$ (Eq. \ref {extremumb}). This can
be thought of as a disorder term $s_{eff} \sim \ln z$ with a
diverging $s_{eff}$. The consequence is an apparent discontinuity,
or even an intrinsic 1st order transition, driven by disorder.

This decoupling transition is consistent with the main features of
the second peak transition: (i) decoupling field being weakly $T$
dependent \cite{Kes,Khaykovich1,Khaykovich2,Deligiannis}, (ii)
decoupling field decreasing with impurity concentration
\cite{Khaykovich1}, (ii) an apparent jump in the critical current
\cite{Kes,Khaykovich1,Khaykovich2,Deligiannis,Giller,Higgins,Marley,Bruynseraede}
(iv) decrease in the c axis critical current \cite{Ooi} and (v) a jump in the Josephson plasma resonance
\cite{Shibauchi,Matsuda2}. The anharmonic region near decoupling
leads to an apparent reduction in $c_{44}$. The reduction in
$c_{44}$ and the resulting reduction in domain sizes account
qualitatively for the enhanced $j_c$. We do not attempt a
quantitative fit; in fact, the measured magnetization changes (and
inferred $j_c$) at the second peak decrease with temperature due
to the strongly temperature dependent relaxation rates
\cite{Yeshurun}, approaching the much smaller equilibrium
magnetizations. 

The nature of the phase at fields above the second peak line has
not been conclusively settled. This work proposes that it is a BG
phase where the domain sizes have been reduced by
$\sqrt{\epsilon}=a^2/(4\pi\sqrt{2}\lam^2)$. Experimentally, the
smooth connection of the second peak with the 1st order line \cite{Avraham} suggests
that it is a single "order-disorder" line of common origin, e.g. a melting line. 
However, the presence of a depinning
line that crosses the "order-disorder" line has been seen by
numerous experiments \cite{Fuchs,Kopelevich,Dewhurst,Ooi}. The crossing of this depinning
line with the "order-disorder" line, separates the latter into a
disorder driven second peak part within a pinned regime and into a
thermally driven part in a depinned, or more weakly pinned regime.
This depinning line corresponds to the onset of a Josephson glass
order, as suggested below.

Consider next the decoupling line at $T>T_0$. This corresponds to
the 1st order transition, which is considered as a melting line
\cite{Kes,Avraham}. However, neutron data \cite{Forgan} has shown
a reentrant behavior in the $600-10^3 G$ range with positional
correlations increasing with temperature. It is possible then that
near the multi-critical point the 1st order line is a decoupling
line. At higher temperatures decoupling then merges into a melting
line.

The 3rd transition line is a transition within the JG order at
$T=T_0$, $B<B_0$ into a weaker JG at $T>T_0$. A depinning line
which is almost vertical at $T\approx T_0$ was indeed observed
\cite{Fuchs, Kopelevich,Dewhurst,Ooi}. We note in particular
the c axis critical current \cite{Ooi} which shows a decrease on
the high temperature side of the depinning line. The thermodynamic 
critical current is proportional to 
the renormalized $z$  that changes from the weakly $T$ dependent
Eq. (51) at $T<T_0$ to the strong exponential decrease with
$T$ in Eq. (53) at $T>T_0$ , consistent with the data. At
$T>T_0$ we also expect a sharp enhancement of the Josephson plasma
resonance, which is an additional tool for identifying the JG
order parameter.

 The final 4th line is a depinning line at $T=B_0T_0/B$, $B>B_0$ 
corresponding to a depinning line as observed in BSCCO
 by current distribution data \cite{Fuchs}, vibration reed \cite{Kopelevich}, 
magnetization \cite{Dewhurst} and c axis
 critical current data \cite{Ooi}. This line is more
difficult to detect by Josephson plasma resonance since its
frequency varies continuously, with discontinuities in derivatives.
 In the decoupled phases (with or without JG order),
where $\langle \cos b \rangle$ is small, we expect the fluctuation
form Eq. (\ref {cosb4}).

We have assumed throughout that our transition lines are well
below melting. Thermal melting is discussed below Eq. (\ref {Td})
while here we estimate the disorder induced melting field. We
assume a Lindeman criterion such that the fluctuations in the
decoupled phase on scale $R^-=a$ are $\langle
u_T^2\rangle=c_L^2a^2$, with $c_L=0.15$ a conventional Lindeman
number$^{21}$. Using Eq. (\ref{R}) with a prefactor as
identified by Eq. (\ref{RBG}) yields a melting field of
$B_m\approx 10^{-2}c_L^3\sqrt{2
B_0B_{\lambda}\lambda_{ab}^5/d\xi_0^4}$ where
$B_{\lambda}=\phi_0/\lambda_{ab}^2$. With BSCCO parameters the
condition $B_m>B_0$ is satisfied if $B_0\lesssim 50B_{\lambda}$,
hence with the second peak field of $B_0\approx B_{\lambda}\approx
500G$ disorder induced melting is expected at a higher field.
 
In conclusion, we have found a phase diagram which is remarkably
close to the experimental one
\cite{Kes,Khaykovich1,Khaykovich2,Deligiannis,Fuchs,Dewhurst,Ooi}, having
a multicritical
point and providing a fundamental
interpretation of both the second peak transition and the more
recently
observed depinning transitions.
 
 \vspace{2mm} 
 
 {\bf Note Added}

In a recent work [H. Beidenkopf, N. Avraham, Y. Myasoedov, H. Shtrikman,
E. Zeldov, and T. Tamegai (unpublished)] the depinning transitions were 
identified by relaxed magnetization data as equilibrium transitions. Both transitions 
at fields below and above the multicritical point were identified and suggested to be 
equilibrium glass transitions.

\vspace{2mm} 
{\bf Acknowledgments}: We thank E. Zeldov, D. T.
Fuchs and P. Le Doussal for most valuable and stimulating
discussions. This research was supported by THE ISRAEL SCIENCE
FOUNDATION founded by the Israel Academy of Sciences and
Humanities.

\newpage
\appendix

\section{Bragg and Josephson glasses}

This section studies nonlinearities due to both disorder and
Josephson coupling leading to two glass order parameters -- the
Josephson glass (JG) and the Bragg glass (BG); the non-singular
phase is neglected. In particular an equivalent term to the
integral $I(z)$ (Eq. \ref {Iz}) is identified and is shown to be
convergent at $k\rightarrow 0$.

 We consider the full Hamiltonian Eq. (\ref {Hfull}), which by
neglecting
 the nonsingular phase becomes
\begin{eqnarray}\label{HbrJ}
&&{\cal
H}/T=\half\sum_{\qq,k,\alpha}(c'\frac{q^4}{k_z^2}+c(k)q^2)|b^{\alpha}(\qq,k)|^2
-\frac{E_J}{T}\sum_{n;\alpha}\int d^{2}r\, \cos
b_{n}^{\alpha}({\bf r}) \nonumber\\
&&- \frac{E_v}{T}\sum_{n;\alpha\neq \beta}\int d^{2}r \cos
[b_{n}^{\alpha}({\bf r}) - b_{n}^{\beta}({\bf r})]
-\frac{g_0}{a^2}\sum_{\alpha\neq \beta,n}\int d^2r \cos [\QQ\cdot
(\uu^{n,\alpha}(\rr)-\uu^{n,\beta}(\rr))]
\end{eqnarray}

The average of the disorder term over the variational Hamiltonian
(\ref {H0}) ${\cal H}_0$ yields
\begin{equation}
\langle \cos [\QQ\cdot (\uu^{n,\alpha}(\rr)-\uu^{n,\beta}(\rr))]
\rangle =\exp \{-\frac{a^2}{2d^2}
\sum_{\qq,k}\frac{q^2}{k_z^2}[G_{\alpha\alpha}(q,k)-G_{\alpha\beta}(q,k)]\}\,.
\end{equation}
We assume for simplicity a square lattice, $Q=2\pi/a$, otherwise a
factor $(aQ/2\pi)^2$ is needed in the exponent; there are then 4
shortest ${\bf Q}$ terms in Eq. (\ref{HbrJ}).  The variational
equation for $G_{\alpha ,\beta}^{-1}(q,k)$, Eq. (\ref
{extremuma}), has now an additional self energy term
$\sigma_{\alpha\beta}^{(1)}$ which allows for an additional RSB.
Written as an equation for matrices in replica space, e.g. ${\hat
G}$, we have
\begin{equation}\label{GA}
{\hat G}^{-1}(q,k)=(c'\frac{q^4}{k_z^2}+c(k)q^2+z){\hat I}-{\hat
\sigma}_2-\frac{q^2}{k_z^2}{\hat \sigma}_1\,.
\end{equation}
When $({\hat \sigma}_1)_{\alpha\beta}=1$, i.e. no RSB, the
previous form (\ref {extremuma}) is recovered. The variational
${\cal F}_{var}$ (\ref {Fvar}) has now a term $\sim \exp [-\half
B_{\alpha\beta}^{(1)}]$ (instead of the $s_0$ term) where
\begin{equation}
B_{\alpha\beta}^{(1)}=\frac{a^2}{d^2}\sum_{\qq,k}\frac{q^2}{k_z^2}
[G_{\alpha\alpha}(q,k)-G_{\alpha\beta}(q,k)]\,.
\end{equation}
The variation of this term identifies
\begin{equation}\label{sigma1}
\sigma_{\alpha\beta}^{(1)}=\frac{4g_0}{d^3}[e^{-\half
B_{\alpha\beta}^{(1)}} -\delta_{\alpha\beta}\sum_{\gamma}e^{-\half
B_{\alpha\gamma}^{(1)}}]
\end{equation}
while $\sigma_{\alpha\beta}^{(2)}$ and $B_{\alpha\beta}^{(2)}$
have the previous forms (\ref {Bab}, \ref{extremumc}). In the
hierarchical scheme $G^{-1}$ is represented by $[{\tilde a},
a(u)]$ which are now given by
\begin{eqnarray}
{\tilde a}&=&c'\frac{q^4}{k_z^2}+c(k)q^2+z-{\tilde \sigma}_2-{\tilde
\sigma}_1\frac{q^2}{k_z^2}\nonumber\\
a(u)&=& -\sigma_2(u)-\sigma_1(u)\frac{q^2}{k_z^2} \,.
\end{eqnarray}
The JG and BG order parameters which measure the degree of RSB are
$\Delta_1(u), \Delta_2(u)$, respectively, where
$\Delta_i(u)=u\sigma_i(u)-\int_0^u\sigma_i(v)dv$, $i=1,2$. Using
the inversion (\ref {inversion}) we can write
\begin{equation}\label{Bi}
\half B_i(u)=\frac{g_i(u)}{u}-\int_u^1 \frac{g_i(v)}{v^2}dv \qquad
i=1,2
\end{equation}
where
\begin{eqnarray}\label{g12}
g_1(u)&=&\frac{a^2}{2d^2}\sum_{\qq,k}[c'q^2+c(k)k_z^2+
(z+\Delta_2(u))\frac{k_z^2}{q^2}+\Delta_1(u)]^{-1}\nonumber\\
g_2(u)&=&\sum_{\qq,k}[c'\frac{q^4}{k_z^2}+c(k)q^2+z+\Delta_2(u)+
\Delta_1(u)\frac{q^2}{k_z^2}]^{-1}\,.
\end{eqnarray}
As in Eq. (\ref {D'}), we find
\begin{equation}\label{Di'}
\frac{\Delta_i'(u)}{u}=-\frac{d}{du}[\frac{\Delta_i'(u)}{g_i'(u)}]
\qquad i=1,2\,.
\end{equation}
Consider first $g_2(u)$ which is dominated by $k\gg q$ so that the
$c'$ term produces just the cutoff $q_u$. The $q$ integration then
yields Eq. (\ref {gu1}) with $\Delta_c\rightarrow
[c(k)+\Delta_1(u)/k_z^2]q_u^2$ in the logarithm. As above, we
replace $k$ by $\pi/d$ in this logarithm since the $k$ integral is
dominated by $k\approx \pi/d$ due to the significant softening of
$c(k)$ near $k=\pi/d$. Hence the form $g_2(u)\sim \ln
[z+\Delta_2(u)]$ is maintained and the solution, as in (\ref {dD})
is a one step function at $u=2t$.

To solve the equation for $\Delta_1(u)$ we simplify the form of
$c(k)$ as
\begin{eqnarray}
c(k)&=&c(0)\equiv c_-  \qquad \qquad k<\frac{1}{a} \nonumber\\
&=& c(\frac{\pi}{d})\frac{4}{d^2k_z^2} \qquad k>\frac{1}{a}
\end{eqnarray}
This form captures the significant dispersion of $c(k)$ with
$c(\pi/d)\ll c(0)$ and allows analytic treatment of the
potentially divergent $k\rightarrow 0$ integrals. The
$k>\frac{1}{a}$ integration range in $g_1(u)$ has an integrand
\[ [c'q^2+\Delta_1(u)+c(\pi/d)(4/d^2)+(z+\Delta_2(u))k_z^2/q^2]^{-1}\]
so that $c(\pi/d)(4/d^2)\gg \Delta_1(u)$ provides a cutoff on the
$q$ integration, i.e. $g_1(u)$ acquires a term independent of
$\Delta_1(u)$. The $k<\frac{1}{a}$ integration has $c(0)/a^2\gg
\Delta_1(u)$ so that after the $k$ integration
\begin{equation}\label{g1}
g_1(u)=\frac{a^2}{8\pi
d^2}\int\frac{q^2dq}{\sqrt{(c'q^2+\Delta_1(u))(c_-q^2+z+\Delta_2(u))}}+const\,.
\end{equation}
$\Delta_1(u)$ varies between $\Delta_1(0)=0$ and $\Delta_1(u_c)$
which depends on the disorder strength (see below);
$\Delta_1(u)=\Delta_1(u_c)$ is constant at $u>u_c$, being a valid
solution of (\ref {Di'}). As the decoupling transition is
approached and $z\rightarrow 0$ the $q$ integration in (\ref {g1})
has distinct forms depending on the ratio of $\Delta_1(u)$ and
$c'z/c_-$. When $\Delta_1(u)<c'z/c_-$ the dominant integration
range is $\Delta_1(u)/c'<q^2<z/c_-$ and the result for the
derivative is
\begin{equation}
\frac{d}{d\Delta_1}g_1(\Delta_1)=\frac{\alpha'}{\sqrt{z}}\ln
\Delta_1   \qquad \alpha'=\frac{a^2}{8\pi d^2c'^{3/2}}\,.
\end{equation}
Substituting in (\ref {Di'}) yields $\Delta_1(u)\approx
u\sqrt{z}/\alpha'\ln ^2u$ and with (\ref {Bi}) we obtain
[$C_2=g_1(u=0)$]
\begin{eqnarray}
g_1(u)&=&C_2+\frac{u}{\ln u} +O(\frac{u}{\ln ^2 u})\nonumber\\
\half B_1(u)&=&C_2+ \ln \ln u +O(\frac{1}{\ln u})
\end{eqnarray}
so that $\sigma_1(u)\sim 1/\ln u \rightarrow 0$ at $u\rightarrow
0$. When $\Delta_1(u)>c'z/c(0)$ the dominant integration range is
$z/c(0)<\Delta_1(u)/c'<q^2$ so that $z=0$ can be taken and
\begin{equation}\label{g1'}
\frac{d}{d\Delta_1}g_1(\Delta_1)=-\frac{\alpha_-}{\sqrt{\Delta_1}}
\qquad \alpha_-=\frac{a^2}{8\pi c'c^{1/2}_-}\,.
\end{equation}
Substituting in (\ref {Di'}) yields $\Delta_1(u)$, so that in both
regimes we have to leading order in $u$
\begin{eqnarray}\label{Du12}
\Delta_1(u)&=& \frac{u\sqrt{z}}{\alpha'\ln ^2 u} \qquad
\Delta_1(u)<z\frac{c'}{c_-}\nonumber\\
 &=& \frac{u^2}{4\alpha_-^2}  \qquad \qquad
\Delta_1(u)>z\frac{c'}{c_-}\,.
 \end{eqnarray}
 Integrating (\ref {g1'}) and using (\ref {Bi}) for $2\alpha_-
 \sqrt{c'z/c_-}<u<u_c$,
 \begin{eqnarray}\label{B1u}
g_1(u)&=&C_2-u\nonumber\\
\half B_1(u)&=&C_2-\ln \frac{u}{u_c}-u_c
\end{eqnarray}
so that $\sigma_1(u)\sim u$ in this range. We suspect that the
solution at $\Delta_1(u)<zc'/c_-$ is significantly modified by the
non-singular phase (as indeed found in Appendix B). This is of no
concern since anyway the effect of this range on the $z$ equation
vanishes (Eq. \ref {A1k} below).

We finally consider the equation for $z$ by using the inversion
formula (\ref {inversion})
\begin{eqnarray}\label{Abg}
A_{\alpha}=\int_{\qq,k}{\tilde
G}(q,k)=&&\int_{\qq,k}\frac{1}{c'\frac{q^4}{k_z^2}+c(k)q^2+z}
[\frac{\sigma_2(0)+\sigma_1(0)\frac{q^2}{k_z^2}}{c'\frac{q^4}{k_z^2}+c(k)q^2+z}+1\nonumber\\
&&+\int_0^1\frac{dv}{v^2}\frac{\Delta_2(v)+\Delta_1(v)\frac{q^2}{k_z^2}}
{c'\frac{q^4}{k_z^2}+c(k)q^2+z+
\Delta_2(v)+\Delta_1(v)\frac{q^2}{k_z^2}}]
\end{eqnarray}
Taking $\sigma_2(0)\sim z$ from section III the $\sigma_2(0)$ term
yields a constant, independent of $z$. Note that without BG order,
$\Delta_1(u)=0$, the one step solution for $\Delta_2(u)$
reproduces the $s_0$ terms in Eq. (\ref {Gtilde}).

Consider first the range $k<1/a$ which led to an apparent
divergence in section III. For small v, where the v integral may
diverge, we take $\Delta_2(v)=0$ so that
\begin{equation}\label{A1A}
A_1=\int_0
\frac{dv}{v^2}\int_{\qq}\int_{k<1/a}k^2[\frac{1}{c'q^4+(c_-q^2+z)k^2}-
\frac{1}{c'q^4+(c_-q^2+z)k^2+\Delta_1(u)q^2}]
\end{equation}
Performing the $k$ integral leads to a $[c_-q^2+z]^{-3/2}$ factor,
which amounts to a lower cutoff $\sqrt{z/c_-}$,
\begin{equation}\label{A1A1}
A_1=\int \frac{dv}{4\pi c^{3/2}_-
v^2}\int_{\sqrt{z/c_-}}dq[-\sqrt{c'}+\frac{1}{q}\sqrt{\Delta_1(v)+c'q^2}]
\end{equation}
For $\Delta_1(v)<zc'/c_-$ one can expand in $\Delta_1(v)$, which
from (\ref {Du12}) yields a term
\begin{equation}\label{A1k}
\int_0^{\sim\sqrt{z}} \frac{dv}{v^2}\frac{v}{\ln ^2 v}\sim
\frac{1}{\ln z}\rightarrow 0
 \qquad z\rightarrow 0 \nonumber \,.
\end{equation}
For the $v$ integration range where $\Delta_1(v)>zc'/c_-$, which
exists if $\Delta_1(u_c)>zc'/c_-$, we have
\begin{equation}\label{A2A}
A_1=\frac{1}{4\pi c^{3/2}_-}[\int_{\sim
\sqrt{z}}\frac{dv}{2v^2}\sqrt{\Delta_1(v)} \ln
\frac{4c_-\Delta_1(v)}{c'z} - \frac{1}{4\alpha_-}\ln z]=
\frac{1}{64\pi \alpha_- c^{3/2}_-}[\ln ^2 z +O(\ln z)]\,.
\end{equation}
The second contribution to $A_{\alpha}$ is from the range $k>1/a$
where $c(k)k_z^2\approx$ constant provides a cutoff in the
$A_{\alpha}$ integrations, hence $\Delta_1(v)$ can be neglected in
the denominator, leading to
\begin{equation}\label{A2s}
A_2=\int_{k>1/a}\int_{\qq}\frac{q^2}{k_z^2[c(k)q^2+z]^2}\int_0\frac{dv}{v^2}\Delta_1(v)
\end{equation}
Identifying $s_0=\int_0\frac{dv}{v^2}\Delta_1(v)$ we obtain the
form (\ref {Iz1}) for $I(z)$, i.e. $A_2=-2s\ln z$. Collecting both
terms we finally have
\begin{equation}\label{ln2}
A_{\alpha}=\frac{\pi d^2\lam^2 }{a^4}\ln^2 z -2s\ln z + O(\ln z)
\end{equation}
where additional $\ln z$ terms involve $\Delta_2(v)$ and $t$ as in
Eqs. ({\ref {A2}, \ref {A3}).

We proceed to identify $\Delta_1(u)$, which determines the BG
domain size, and to examine the condition $\Delta_1(u_c)>zc'/c_-$
necessary for the appearance of the $\ln^2z$ term in (\ref {ln2}).
Eqs. (\ref {sigma1},  \ref {B1u}) yield for the range $2\alpha
 \sqrt{c'z/c_-}<u<u_c$,
\begin{equation}\label{sigma11}
\sigma_1(u)=\frac{4g_0}{d^3}\frac{u}{u_c}e^{-C_2+u_c}\,.
\end{equation}
The definition $\Delta_1(u)=u\sigma_1(u)-\int_0^u\sigma_1(v)dv$
then leads to
\begin{equation}\label{D11}
\Delta_1(u_c)=\frac{2g_0}{d^3}u_c e^{-C_2+u_c}
\end{equation}
$C_2$ is a Debye Waller factor which is small by the assumption of
being well below melting, $T/\tau\ll 1$. Comparing with (\ref
{Du12}) we identify $u_c\approx 10^4sTd^2\xi_0^4/\lam^2a^4\ll 1$
and $\Delta_1(u_c)\approx \alpha_-(g/d^3)^2$.

$\Delta_1(u_c)$ is related to the BG domain size in the axis
perpendicular to the layers $L_{BG}^-=\sqrt{c_-/\Delta_1(u_c)}$ or
in the ab plane $R_{BG}^-=\sqrt{c'/\Delta_1(u_c)}$, as identified
by the $q,k$ cutoffs in $g_1(u)$ (Eq. \ref {g12}), or by
evaluating displacement correlations \cite{Giamarchi}. Hence
\begin{equation}\label{RBG}
R_{BG}^-\approx \frac{10^{-4}\lam^3 a^3}{sd\xi_0^4}
\end{equation}
while $L_{BG}^-=R_{BG}^-a/(\lam\sqrt{2\pi})$. These forms are
valid close to decoupling [$\Delta_1(u_c)>zc'/c_-$] or in the
decoupled phase ($z=0$). Remarkably, this result of $R_{BG}$ is,
up to the $10^{-4}$ factor, identical to that found from the
dimensonal analysis Eq. (\ref {R}) with $\langle u_T^2 \rangle
\approx a^2$. We do not attempt to evaluate $R_{BG}$ in the
coupled phase with $\Delta_1(u_c)<zc'/c_-$ since then the
nonsingular phase, being neglected here, is essential for
generating the proper $c_{44}$. As noted above, for the purpose of
decoupling the value of $A_{\alpha}$ in the $k<1/a$ range for
large $z$ [$zc'/c_->\Delta_1(u_c)$] is negligible even without the
nonsingular phase, as seen in (\ref {A1k}) .

The condition $\Delta_1(u_c)>zc'/c_-$, for the appearance of the
$\ln ^2z$ in Eq. (\ref {ln2}) can be written in terms of
$\lambda_c^R$ (Eq. \ref {lamcR}) with
$\sqrt{z/c_-}=1/\lambda_c^R\sqrt{\epsilon}$,
\begin{equation}\label{cond2}
\lambda_c^R > R_{BG}^-/\sqrt{\epsilon}\approx
\frac{10^{-3}a\lam^5}{sd\xi_0^4}\,.
\end{equation}
For typical BSCCO or YBCO parameters this implies a renormalized
anisotropy of $\lambda_c^R/\lam>10^5$, i.e. fairly close to
decoupling at $z=0$. Note that $R_{BG}^-/\sqrt{\epsilon}$ can be
identified as $R_{BG}^+$, the BG domain size in the coupled phase,
as shown in Appendix B and section IV.

\section{Bragg Glass with non-singular phase}
We solve here the decoupling transition with nonlinear coupling of
disorder (BG effects) and with the non-singular phase. The $E_v$
term of Eq. (\ref {Hfull}) is neglected, i.e. no JG effects. This
describes correctly thermal decoupling, i.e. the line $s+t=1$ in
Fig. 1 where JG is absent within the RSB scheme. To identify the
proper ${\cal H}_0$, we expand the renormalized Josephson coupling
$-z\cos [b_n^{\alpha}(\rr)+\theta_n^{\alpha}(\rr)]\approx \half z
[b_n^{\alpha}(\rr)+\theta_n^{\alpha}(\rr)]^2$ so that with the
other Gaussian terms of (\ref {Hpure2}) we have
\begin{eqnarray}\label{H03}
{\cal H}_0=\half \int
\frac{d^2qdk}{(2\pi)^3}[G_f^{-1}(q,k)|\theta^{\alpha}(q,k)|^2&&+
z|\theta^{\alpha}(q,k)+b^{\alpha}(q,k)|^2+c(q,k)q^2|b^{\alpha}(q,k)|^2\nonumber\\
&&-\frac{q^2}{k_z^2}\sigma_{ab}b^{\alpha *}(q,k)b^{\beta}(q,k)]\,.
\end{eqnarray}
Formally, one needs to perform a variation of $\langle \cos
[b_n^{\alpha}(\rr)+\theta_n^{\alpha}(\rr)]\rangle=\exp [-\half
A_{\alpha}]$ where
\begin{equation}\label{AB}
A_{\alpha}=\sum_{\qq,k}\langle|\theta^{\alpha}(q,k)+b^{\alpha}(q,k)|^2\rangle
\end{equation}
to obtain the $z$ term in (\ref {H03}). This procedure was also
used for decoupling in presence of columnar defects
\cite{Morozov}. We proceed as in the pure case (\ref {Hpure3}) by
shifting to
\begin{equation}\label{thetat}
{\tilde
\theta}^{\alpha}(q,k)=\theta^{\alpha}(q,k)+\frac{z}{G_f^{-1}(q,k)+z}b^{\alpha}(q,k)
\end{equation}
which yields
\begin{eqnarray}\label{H04}
{\cal H}_0&=&\half \int [(G_f^{-1}(q,k)+z)|{\tilde
\theta}^{\alpha}(q,k)|^2+G_{\alpha\beta}^{-1}(q,k)b^{\alpha
*}(q,k)b^{\beta}(q,k)]\nonumber\\
G_{\alpha\beta}^{-1}(q,k)&=&[c'\frac{q^4}{k_z^2}+c(k)q^2+
\frac{zq^2}{q^2+(1+\lam^2k_z^2)/(\lambda_c^R)^2}]\delta_{\alpha\beta}
-\frac{q^2}{k_z^2}\sigma_{\alpha \beta}
\end{eqnarray}
where the last term corresponds to the $B^2$ term of $c_{44}$ (Eq.
\ref {c44}) with $\lambda_c$ replaced by $\lambda_c^R$. Note that
for $q\gg (1+\lam^2k_z^2)^{1/2}/\lambda_c^R$ this reduces to (\ref
{GA}) with $\sigma_2\rightarrow 0$. A term corresponding to the
last term of (\ref {c44}), being $\sim (\lambda_c^R)^{-2}$, is
neglected.

We proceed to evaluate $g_1(u)$ with (\ref {g12}) replaced here by
\begin{equation}\label{g1B}
g_1(u)=\frac{a^2}{2d^2}\sum_{\qq,k}[c'q^2+[c(k)+\frac{z}
{q^2+(1+\lam^2k_z^2)/(\lambda_c^R)^2}]k_z^2 +\Delta_1(u)]^{-1}\,.
\end{equation}
For $k>1/a$ $c(k)k_z^2\gg \Delta_1(u)$ and $g_1(u)$ is $\Delta_1$
independent, as in Appendix A. For $k<1/a$ two regimes are
identified, where the coefficient of the $k_z^2$ term in (\ref
{g1B}) becomes
\begin{eqnarray}
c_+=c(0)+z(\lambda_c^R)^2=c_-+\frac{\lam^2\tau}{4\pi Td^3} \qquad
q<1/\lambda_c^R \nonumber\\
c_-=c(0)=\frac{a^4\tau}{2(4\pi d)^3\lam^2 T} \qquad
q>1/\lambda_c^R
\end{eqnarray}
so that $c_+/c_-=1+1/\epsilon \gg 1$ with $\epsilon$ defined in
(\ref {epsilon}). This reflects the significant dependence of
$c_{44}$ on interchanging the $q\rightarrow 0$ and
$1/\lambda_c^R\rightarrow 0$ limits, as discussed in section IV.
After the k integration we obtain (replacing \ref {g1})
\begin{equation}\label{g1C}
g_1'(\Delta_1)=-\frac{a^2}{8\pi
d^2}[\int_0^{1/\lambda_c^R}\frac{1}{\sqrt{c_+}}+\int_{1/\lambda_c^R}
\frac{1}{\sqrt{c_-}}]\frac{qdq}{\sqrt{c'q^2+\Delta_1}}\approx
\alpha_{\pm}/\sqrt{\Delta_1}
\end{equation}
where $\alpha_{\pm}=a^2/(8\pi c'\sqrt{c_{\pm}})$ with $\alpha_+$
for $\sqrt{\Delta_1(u)/c'}>1/\lambda_c^R$ and $\alpha_-$ for
$\sqrt{\Delta_1(u)/c'}<1/\lambda_c^R$. Hence
$\Delta_1(u)=u^2/4\alpha_{\pm}$ and Eqs. (\ref {B1u}, \ref
{sigma11}) are valid in both $\alpha_{\pm}$ regimes. Comparing
(\ref {D11}) with $u^2/4\alpha_{\pm}$ identifies $u_c\approx
2\alpha_{\pm}g_0/d^3$ and $\Delta_1(u_c)\approx
\alpha_{\pm}(g_0/d^3)^2$. The BG scales
$R_{BG}^{\pm}=\sqrt{c'/\Delta_1(u_c)}$ are therefore
\begin{eqnarray}\label{R+-}
R_{BG}^+&&\approx  \frac{10^{-3}a\lam^5}{sd\xi_0^2}\qquad
R_{BG}^+<1/\lambda_c^R\nonumber\\
R_{BG}^-&&\approx \frac{10^{-4}\lam^3 a^3}{sd\xi_0^4}\qquad
R_{BG}^->1/\lambda_c^R
\end{eqnarray}
so that
$R_{BG}^+=\sqrt{\alpha_+/\alpha_-}R_{BG}^-=R_{BG}^-/\sqrt{\epsilon}$.
The range $R_{BG}^-<\lambda_c^R<R_{BG}^+$ allows for both length
scales and serves as a crossover between the regimes in (\ref
{R+-}). The ratio $R_{BG}^+=R_{BG}^-/\sqrt{\epsilon}$ reflects the
change in elastic constants, as in the dimensional argument of
section IV. The result  (\ref {R+-}) for $R_{BG}^-$ agrees with
(\ref {RBG}) in Appendix A.

Renormalization of $z$ requires the sum (\ref {AB}) which is
averaged with respect to ${\cal H}_0$ of (\ref {H04})
\begin{eqnarray}\label{AC}
A_{\alpha}&&=\sum_{\qq,k}|{\tilde
\theta}^{\alpha}(q,k)-\frac{z}{G_f^{-1}(q,k)+z}b^{\alpha}(q,k)+
b^{\alpha}(q,k)|^2\rangle\nonumber\\
&&=\sum_{\qq,k}[\frac{1}{G_f^{-1}(q,k)+z}+(\frac{z}{G_f^{-1}(q,k)+z})^2G_{\alpha
\alpha}(q,k)] \,.
\end{eqnarray}
The first term is $\approx (T/\tau)\ln z$ and is neglected at
$T\ll \tau$. The second term has a factor
\begin{equation}
\frac{z}{G_f^{-1}(q,k)+z}=\frac{q^2}{q^2+(1+\lam^2k_z^2)/(\lambda_c^R)^2}
\end{equation}
which for $q<(1+\lam^2k_z^2)^{1/2}/\lambda_c^R$ strongly reduces
the $q$ integration, while for larger $q$, $A_{\alpha}$ becomes
\begin{equation}\label{AB2}
A_{\alpha}=\int_{\qq,k}^{'}
\frac{1}{c'\frac{q^4}{k_z^2}+c(k)q^2+z}[1+\int_0^1\frac{dv}{v^2}
\frac{\Delta_1(v)\frac{q^2}{k_z^2}}
{c'\frac{q^4}{k_z^2}+c(k)q^2+z+\Delta_1(v)\frac{q^2}{k_z^2}}]
\end{equation}
where $\int^{'}$ indicates $q>(1+\lam^2k_z^2)^{1/2}/\lambda_c^R$.
For $k>1/a$, $c(k)k_z^2\gg \Delta_1(u)$ provides a cutoff with the
result $A_2=2s\ln (\Delta_c/z)$ as in Eq. (\ref {A2s}). For
$k<1/a$ the $v$ integral term of (\ref {AB2}) becomes $A_1$ as in
(\ref {A1A}) except for a $q$ cutoff in $\int^{'}$. The $k$
integration of (\ref {A1A}) produces a cutoff
$q>\sqrt{z/c_-}=1/(\lambda_c^R\sqrt{\epsilon})\gg
(1+\lam^2k_z^2)^{1/2}/\lambda_c^R$, hence (\ref {A1A1}) is valid.
For $\Delta_1(v)<zc'/c_-$
\begin{equation}
\frac{1}{\sqrt{z}}\int_0^{\sim
\sqrt{z}}\frac{dv}{v^2}\Delta_1(v)\sim \mbox{const.}
\end{equation}
while for $\Delta_1(v)>zc'/c_-$  (\ref {A2A}) is reproduced. The
latter integration range exists if $\Delta_1(u_c)>zc'/c_-$, i.e.
$R_{BG}^-<\sqrt{c_-/z}=\lambda_c^R\sqrt{\epsilon}$. Using (\ref
{R+-}) we identify the condition for the appearance of the
$\ln^2z$ term as
\begin{equation}\label{cond}
R_{BG}^+< \lambda_c^R  \qquad \qquad  \textrm{onset of $\ln^2 z$
term}\,.
\end{equation}
This is also the condition found in Appendix A (Eq. \ref {cond2}),
as well as the condition of section IV, as illustrated in Fig. 3,
for the onset of the anharmonic regime.

\section{Josephson glass with non-singular phase}

In this appendix we extend the solution of section III to include
the nonsingular phase. In particular we identify the pinning
length $R_p$ in the coupled phase and show that it coincides with
(\ref {R}) (with $\langle u_T^2 \rangle \approx \xi_0^2$), up to a
numerical prefactor. Since disorder is linearized, we do not
expect to derive BG domain sizes. Also the integral $I(z)$ is
reconsidered.

Consider then Eq. (\ref {Hreplica}) with the pure part replaced by
Eq. (\ref {Hpure2}). The harmonic part can be written as
 \begin{eqnarray}
 &&G_f^{-1}(q,k)|{\tilde b}^{\alpha}(q,k)-b^{\alpha}(q,k)|^2+
[c(q,k)q^2\delta_{\alpha\beta}-s_0\frac{q^2}{k_z^2}]b^{\alpha}(q,k)b^{\beta* }(q,k)
\nonumber\\
 &&=G_f^{-1}(q,k)|{\tilde b}^{\alpha}(q,k)|^2 +
 d^{\alpha}B^{-1}_{\alpha\beta}(q,k)d^{\beta *}(q,k)-G_f^{-2}(q,k)
 B_{\gamma\alpha}{\tilde b}^{\gamma}(q,k){\tilde b}^{\alpha *}(q,k)
\end{eqnarray}
where
\begin{eqnarray}\label{shift}
d^{\alpha}(\qq,k)&=&b^{\alpha}({\bf q},k)-B_{\gamma,\alpha}(q,k)
G_{f}^{-1}(q,k)\tilde{b}^{\alpha}(\qq,k)\nonumber\\
B^{-1}_{\alpha,\beta}(q,k)&=&G_{f}^{-1}(q,k)\alpha(q,k)
\delta_{\alpha,\beta} - s_0\frac{q^2}{k_z^2}\nonumber\\
\alpha(q,k)&=&1+G_f(q,k)c(q,k)q^2  \,.
\end{eqnarray}
The resulting replicated Hamiltonian is
\begin{eqnarray}\label{Hr1}
{\cal H}^{(2)}/T = &&\half \sum_{\qq,k;\alpha,\beta}
B_{\alpha,\beta}^{-1} d^{\alpha}({\bf q},k)d^{\beta *}({\bf q},k)
+\half [\frac{c(q,k)}{\alpha(q,k)}q^2\delta_{\alpha,\beta} -
\frac{s_0q^2}{\alpha^2(q,k)k_z^2}]\tilde{b}^{\alpha}({\bf q},k)
\tilde{b}^{\beta *}({\bf q},k)\nonumber\\
&&-\frac{E_J}{T}\sum_{n;\alpha}\int d^2r\cos
\tilde{b}_n^{\alpha}({\bf r}) -\frac{E_v}{T}\sum_{n;\alpha\neq
\beta}\int d^2r\cos [\tilde{b}_n^{\alpha}({\bf
r})-\tilde{b}_n^{\beta}({\bf r})]\,.
\end{eqnarray}

The effect of the nonsingular phase on our previous Hamiltonian
Eq. (\ref {Hreplica}) of section II is to replace
$c(q,k)\rightarrow c(q,k)/\alpha(q,k)$ and $s_0\rightarrow
s_0/\alpha^2(q,k)$. From the definition in Eq. (\ref {shift}) we
find that $\alpha(q,k)-1$ is either $\sim q^2$ or $\sim k^2$ and
is small except when
\begin{eqnarray}\label{alph}
\alpha(q,k)-1&=&\epsilon \qquad k<1/\lambda_{ab}, \,
q<ka/\lambda_{ab}\nonumber\\
&=& \frac{a^2}{16\pi\lambda_{ab}^2} \qquad k<1/\lambda_{ab},
\,q>ka/\lambda_{ab}
\end{eqnarray}
This behavior is sufficient to eliminates the $k\rightarrow 0$
divergence of $I(z)$ (leading to $\sim 1/\sqrt{z}$) as shown
below.

We proceed to evaluate the fluctuations in $u^{tr}({\bf q},k)$ and
identify the scale $R_p$. From eq. (\ref {shift})
\begin{equation}
\langle |u_T({\bf q},k)|^2\rangle=(2\pi
d^2)^{-2}\frac{q^2}{k_z^2}[\langle
d^{\alpha}(\qq,k)d^{\alpha*}(\qq,k)\rangle
+G_f^{-2}(q,k)B_{\gamma\alpha}(q,k)B_{\gamma'\alpha}(q,k)G_{\alpha\beta}(q,k)]\,.
\end{equation}
Here $G_{\alpha\beta}(q,k)= \langle {\tilde
b}_{\gamma}(\qq,k){\tilde b}_{\gamma'}^*(\qq,k)\rangle$ is the
solution from section III, and in the replica limit
\begin{eqnarray}
\sum_{\gamma\gamma'}G_{\gamma\gamma'}(q,k)&&\rightarrow  0 \nonumber\\
\sum_{\gamma}G_{\alpha\gamma}(q,k)&& \rightarrow
[\frac{c(q,k)}{\alpha(q,k)}q^2+z]^{-1}
\end{eqnarray}
where terms involving $\Delta_0$ cancel. Hence
\begin{eqnarray}
\langle |u_T({\bf q},k)|^2\rangle=&&(2\pi
d^2)^{-2}\frac{q^2}{k_z^2}
[B_{\alpha\alpha}(q,k)+\frac{G_f^{-2}(q,k)}{(G_f^{-1}(q,k)+c(q,k)q^2)^2}
{\tilde G}(q,k)\nonumber\\
&&+\frac{2s_0q^2/k_z^2}{(G_f^{-1}(q,k)+c(q,k)q^2)^3}
\frac{G_f^{-2}(q,k)}{\frac{c(q,k)}{\alpha(q,k)}q^2+z}] \,.
\end{eqnarray}
With some straightforward algebra,
\begin{eqnarray}
\langle |u_T({\bf q},k)|^2\rangle &=& (2\pi
d^2)^{-2}\frac{q^2}{k_z^4} [s_0q^2 G_f({\bf q},k)\alpha^{-1}({\bf
q},k) \left(c({\bf q},k) q^2 + \frac{G_f^{-1}({\bf
q},k)z}{G_f^{-1}({\bf
q},k)+z}\right)^{-1}\nonumber\\
&+& \frac{s_0}{c({\bf q},k)\alpha^2({\bf q},k)}\left(\frac{c({\bf
q},k)} {\alpha({\bf q},k)}q^2 + z\right)^{-1}] + \ldots
\label{fluc}
\end{eqnarray}
where $\ldots$ stands for terms which converge in $({\bf q},k)$
integration. Note the term $G_f^{-1}({\bf q},k)z/[G_f^{-1}({\bf
q},k)+z]$ which depends on the order of $q\rightarrow 0$ and
$z\rightarrow 0$ limits; this limit dependence leads to the
apparent discontinuity in $c_{44}$ as discussed in section IV. For
$z\neq 0$ and small $q$, i.e. $G_f^{-1}({\bf q},k) \ll z$ the
first term in Eq.\ (\ref{fluc}) dominates, leading to
\begin{equation}
\langle |u_T({\bf q},k)|^2\rangle \approx \frac{4\pi^2s_0T^2}
{a^8[c_{44}k^2 + c_{66}q^2]^2}  \hspace{10mm} q<1/\lambda_c^R
\end{equation}
where $c_{44}$ is from Eq.\ (\ref{c44a}) and the condition
$G_f^{-1}({\bf q},k) \ll z$ is written in terms of $\lambda_c^R$
(Eq. \ref {lamcR}). The correlations at distance $r$ parallel to
the layers are then
\begin{equation}
\langle [u_T(r)-u_T(0)]^2\rangle \approx
\frac{4d^2s_0T^2}{a^4c_{44}^{1/2}c_{66}^{3/2}}r \equiv
\xi_0^2\frac{r}{R_p} \label{Rp1}\, .
\end{equation}
The last equality defines the pinning length $R_p$ where the
fluctuations become of order $\xi_0^2$. This result for $R_p$ (up
to a numerical prefactor) is
 the same as the one obtained from Eq.\ (\ref{R}) with
 $\langle u_T^2 \rangle \approx \xi_0^2$.

In the decoupled phase with $z=0$ the second term in Eq.\
(\ref{fluc}) dominates. To leading order in $\epsilon$ the result
is identical to Eq.\ (\ref{Rp1}) except that $c_{44}$ is replaced
by its $z=0$ value Eq.\ (\ref{c44b}), i.e. the pinning length is
reduced.

Consider next the integral $I(z)$. As noted below Eq. (\ref {Hr1})
the nonsingular phase leads to the replacements $c(q,k)\rightarrow
c(q,k)/\alpha(q,k)$ and $s_0\rightarrow s_0/\alpha^2(q,k)$ so that
Eq. (\ref {Iz}) becomes
\begin{equation}\label{Iz2}
I(z)=\int\frac{dq^2dk}{k_z^2c(q,k)\alpha(q,k)}\frac{1}{[c(q,k)/
\alpha(q,k)]q^2+z} \,.
\end{equation}
In the range $1/a<k<\pi/d$ with $\alpha(q,k)\approx 1$ the $q^2$
term in $c(q,k)$ amounts to a cutoff $q_u^2$ (defined below (\ref
{C1})) leading to $I_0(z)$ (\ref {Iz3}). In the range
$1/\lam<k<1/a$ we have $c(q,k)=c(0)(1+2\pi \lam^2q^2/a^2k^2)$ and
$\alpha(q,k)\approx 1$. The singularity in $z$ which we wish to
identify, is exhibited by $q\rightarrow 0$, hence $c(q,k)\approx
c(0)=c_-$ leads to the first correction
\begin{equation}
I_1(z)=-\frac{2\lam}{c^2_-}\ln z +const.
\end{equation}
In the range $k<1/\lam$ we have two terms
\begin{eqnarray}
I_2(z)&=&2\int_0^{1/\lam}\frac{dk}{k^2c_-} \int_0^{\lam
k/a}\frac{dq^2}{[1+\frac{2\pi\lam^2q^2}{a^2k^2}]
[c_-q^2(1+\frac{2\pi\lam^2q^2}{a^2k^2})+z]}\nonumber\\
I_3(z)&=&2\int_0^{1/\lam}\frac{dk}{k^2c''} \int_{\lam
k/a}^{1/a}(\frac{16\pi\lam^2k^2}{a^2q^2})^2 \frac{dq^2}{c''q^2+z}
\end{eqnarray}
where $c''=(32\pi^2\lam^4/a^4)c_-$ is due to the finite effect of
$c(q,k)/\alpha(q,k)$ when $q>\lam k/a,\,k<1/\lam$. In $I_2(z)$ the
$q^4$ term replaces the $q^2$ cutoff as $ak/\lam$ leading to
\begin{equation}\label{I4}
I_2(z)=\frac{a}{\lam^2c^2_-}\sqrt{\frac{c_-}{z}} +
\frac{2\lam}{c^2_-}\ln z  +const.
\end{equation}
while
\begin{equation}
I_3(z)=\frac{(8\pi)^3\lam}{3ac''^2}\sqrt{\frac{c''}{z}}
\end{equation}
is smaller then the 1st term of (\ref {I4}). We conclude then
\begin{equation}
I(z)=\frac{a}{\lam^2c^2_-}\sqrt{\frac{c(0)}{z}} +2s\ln
\frac{\Delta_c}{z} + const.
\end{equation}

The effect of the $1/\sqrt{z}$ term is significant, in terms of
$\lambda_c^R$ (Eq. \ref {lamcR})  if
\begin{equation}
\frac{\lambda_c^R}{\lam}> \frac{\sqrt{2}\lam^2a}{4d^2\ln
^2(a/d)}\approx 10^4
\end{equation}
for BSCCO parameters; with bare anisotropy of $\lambda_c/\lam
\approx 50$ one needs to be fairly close to the transition to have
an effect from the $1/\sqrt{z}$ term. Note that nonlinear coupling
of disorder, i.e. BG formulation, is much more efficient in
reducing the $I(z)$ singularity, as shown in Appendices A and B.

\end{document}